\documentclass[12pt]{article}
\usepackage{array}
\usepackage{amsthm}
 \usepackage{a4wide}
\usepackage{bibunits}
\defaultbibliographystyle{unsrt}
\usepackage[utf8]{inputenc}
\usepackage[colorlinks=true,linkcolor=magenta,citecolor=blue,urlcolor=blue]{hyperref}
\usepackage{physics}
\usepackage{newtxtext}
\usepackage{here}
\usepackage{mathtools}
\usepackage{bm}
\usepackage{fancybox}
\usepackage[export]{adjustbox}
\usepackage{authblk}
\usepackage{stmaryrd}
\usepackage[T1]{fontenc}
\usepackage[english]{babel}
\usepackage{color}
\usepackage{comment}
\usepackage{amsmath,amssymb,amsfonts,mathrsfs}

\usepackage{graphicx}
\newcommand{\mc}[1]{\mathcal{#1}}
\newcommand{\mr}[1]{\mathrm{#1}}

\newcommand{\mb}[1]{\mathbb{#1}}
\newcommand{\msf}[1]{\mathsf{#1}}
\newcommand{\mf}[1]{\mathfrak{#1}}

\newcommand{\blue}[1]{\textcolor{blue}{#1}}

\usepackage{tikz}

\newcommand{\plaquetteconfig}[4]{%
\begin{tikzpicture}[scale=0.4,every node/.style={font=\small},baseline={(0.,0.1)}]
  \node (A) at (0,0) {$#3$};
  \node (B) at (1,0) {$#4$};
  \node (C) at (1,1) {$#2$};
  \node (D) at (0,1) {$#1$};
\end{tikzpicture}%
}

\newcommand{\plaquetteconfigwide}[4]{%
\begin{tikzpicture}[scale=0.4,every node/.style={font=\small},baseline={(0.,0.1)}]
  \node (A) at (0,0) {$#3$};
  \node (B) at (1.3,0) {$#4$};
  \node (C) at (1.3,1.3) {$#2$};
  \node (D) at (0,1.3) {$#1$};
\end{tikzpicture}%
}

\theoremstyle{remark}

\begin{document}
\title{Non-Hermitian Quantum Many-Body Scar Phase}
\author[1,2,3]{Keita Omiya}
\author[3]{Yuya O. Nakagawa}
\affil[1]{PSI Center for Scientific Computing, Theory and Data, 5232 Villigen PSI, Switzerland}
\affil[2]{Institute of Physics, Ecole Polytechnique F\'ed\'erale de Lausanne (EPFL), CH-1015 Lausanne, Switzerland}
\affil[3]{QunaSys Inc., Aqua Hakusan Building 9F, 1-13-7 Hakusan, Bunkyo, Tokyo 113-0001, Japan}
\date{}
\maketitle

\begin{abstract}
		We introduce a novel non-equilibrium phase---the quantum many-body scar (QMBS) phase---that emerges in non-Hermitian many-body dynamics when scarred wavefunctions are selectively stabilized via non-Hermitian driving. Projective measurements, or non-Hermitian counterparts, preferentially reinforce QMBS, counteracting the entropy growth that drives thermalization. As a result, atypical, high-energy scarred wavefunctions that are negligible in the long-time dynamics of closed systems become non-equilibrium steady states. We establish the existence of the QMBS phase and its sharp, first-order phase transition from an ergodic thermal phase, through both analytical arguments and numerical simulations of three representative models: a random quantum circuit model, the $SU(q)$ spin model, and the paradigmatic spin-1 XY model.     
\end{abstract}

\section{Introduction}
Thermalization is a ubiquitous phenomenon in generic many-body systems. Under time evolution, memory of an initial state will eventually be erased, and the steady state will be described by the Gibbs ensemble. Elucidating this process has long been a central challenge in statistical physics. In quantum mechanics, the linearity of the Schr\"odinger equation guarantees the invertibility of the time-evolution, which formally preserves all information about the initial state within the wavefunction at any instance. 

Recent years have witnessed a remarkable progress on demystifying this apparent paradox, mainly through the seminal works by Deutsch and Srednicki~\cite{eth_deutsch,eth_srednicki}, which introduced the Eigenstate Thermalization Hypothesis (ETH). The ETH posits that an individual eigenstate already exhibits thermal properties: the quantum mechanical expectation values of macroscopic observables in a single eigenstate coincide with the statistical expectation value, up to small fluctuations. This paradigm has fundamentally advanced our understanding of quantum thermalization, leading to the notion of typicality~\cite{PhysRevLett.80.1373,PhysRevLett.96.050403,sugita2006}. The ETH is now regarded as the standard framework. 

While the ETH is believed to hold for generic quantum many-body systems, several exceptional non-thermal systems violating its prediction have been identified. A prominent example is integrable systems: arising from the fine-tuned structure of interactions, they have an extensive number of local conserved quantities, which lead to corrections to the Gibbs ensemble (generalized Gibbs ensemble)~\cite{PhysRevLett.98.050405}. Another example is the many-body localization (MBL), where strong disorder prohibit spread of quantum information and entanglement~\cite{Basko_2006}. Similar to the integrable systems, the existence of an extensive number of (quasi-)local conserved quantities, local integral of motions (LIOMs), break ergodicity~\cite{PhysRevLett.111.127201,PhysRevB.90.174202}. 

For integrable and many-body localized systems, local conserved quantities are responsible for the systems' non-thermal behavior. Therefore, one naturally expects that the absence of such conservation laws would imply the ETH. However, this speculation was disproved by the seminal work by Shiraishi and Mori~\cite{PhysRevLett.119.030601}, which discusses a systematic recipe to construct non-thermal eigenstates in non-integrable spin models without disorder. Almost simultaneously, an experiment using a Rydberg atom chain observed the possible dynamical signature of such non-thermal eigenstates~\cite{Bernien_2017}. These eigenstates were dubbed quantum many-body scar (QMBS) states or scar states for short~\cite{Turner_2018}, due to their similarity to the (single-particle) quantum scar in chaotic billiards~\cite{PhysRevLett.53.1515}.

A distinct characteristic of QMBS, in contrast to the integrability and the MBL, is that their non-thermal behavior does \textit{not} stem from (conventional) symmetries (note Ref.\,\cite{PhysRevX.14.041069} for an attempt to include QMBS as a ``generalized'' symmetry), but rather emerges solely from non-thermal eigenstates. This very feature, however, also limits their broader physical interpretation: unless an initial state has an appreciable overlap with one of the scar states as in the Rydberg experiment, the contribution of scar states to the system's long time dynamics is negligible (for a short time, some implications of QMBS may be visible even for generic initial states, see Refs.\,\cite{PhysRevX.13.011033,b83y-h128}), since their weight in the many-body Hilbert space vanishes in the thermodynamic limit. Moreover, given the extreme sensitivity of many-body eigenstates against perturbations (e.g., orthogonality catastrophe~\cite{PhysRevLett.18.1049}), even infinitesimal disorders may disrupt the structure of the Hamiltonian that supports scar states, therefore restoring the conventional thermalization. The problem of the stability of QMBS has been discussed for a long time, and some numerical results have suggested some sort of robustness~\cite{PhysRevResearch.2.033044,PhysRevB.103.104302,PRXQuantum.4.040348}. However, these results have focused on the early time behavior in the quench dynamics and the eventual thermalization seems unavoidable. 

Situation may be different when we consider instead open quantum systems whose dynamics is not generated by Hermitian Hamiltonian. For example, Ref.\,\cite{PhysRevLett.133.216601} points out that adding non-Hermitian terms can stabilize the scarred dynamics of the PXP chain, stemming from the ``Fock skin'' effect. Furthermore, QMBS possess some interesting connections with non-Hermitian physics even at the mathematical level: the scar states of the PXP model becomes exactly solvable when a non-Hermitian term is added, and non-Hermitian terms naturally emerge in several well-known QMBS models including Affleck-Kennedy-Lieb-Tasaki (AKLT) model~\cite{PhysRevA.107.023318, PhysRevB.108.054412}. This implies that QMBS may be intricately related to non-Hermitian physics, and physical implications are thus visible in this regime. 

In this paper, we establish QMBS as a unique non-equilibrium \textit{phase} of matter under non-Hermitian dynamics, which we dub QMBS phase. Similar to the measurement induced phase transition (MITP)~\cite{PhysRevB.98.205136,PhysRevB.99.224307,PhysRevX.9.031009}, the QMBS phase is a distinctive phase peculiar to non-Hermitian physics, which is induced by the competition between entropy production by the thermal state and the selective stabilization of scarred wavefunction by non-Hermitian driving. 
Notably, the QMBS phase may have the volume law entanglement entropy unlike MITP, but it is still distinct from thermalization. In the QMBS phase, scarred wavefunction is robust against perturbations in a genuine sense, which opens up a pathway to observation and manipulation of QMBS in the laboratory.

The rest of the paper is organized as follows. In Sec.\,\ref{sec:motivation}, we provide motivation for the subsequent discussions, recalling the basic algebraic structure of QMBS. Particularly, we demonstrate that a certain non-Hermitian term can stabilize non-thermalizing dynamics arising from QMBS. In Sec.\,\ref{sec:rqc}, we analyze a one-dimensional random quantum circuit model that supports the QMBS structure introduced in Sec.\,\ref{sec:motivation}. The dynamics of this circuit model can be mapped to a two-dimensional classical statistical mechanical model, and all the numerical methods (the mean-field approximation, the density matrix renormalization group, and the real space renormalization group) consistently demonstrate the existence of a first-order phase transition and the corresponding QMBS phase. In Sec.\,\ref{sec:eft}, we further investigate the nature of this transition by developing an effective semiclassical theory within the Ginzburg-Landau paradigm. In the limit which justifies the mean-field approximation, the effective degrees of freedom turns out to be $SU(1,1)$ spins, and verify the first-order phase transition. We also find that the leading correction to the mean-field ansatz is strongly dependent on the underlying phase, which modifies the phase diagram. We present a exact diagonalization study on a QMBS Hamiltonian, which exhibits a sharp transition from the thermal to the QMBS phase. We conclude our discussion in Sec.\,\ref{sec:conclusion}.

\section{Motivation: stabilizing non-thermal dynamics}
\label{sec:motivation}
In this preliminary section, we briefly introduce a key idea of a systematic recipe for QMBS Hamiltonians known as the Shiraishi-Mori (SM) construction. We observe that non-Hermitian terms naturally arise if the SM construction is applied to Hermitian models with some ``non-trivial'' QMBS models, suggesting connections between QMBS and non-Hermitian physics at the mathematical level. To demonstrate their physical significance, we simulate a simple non-Hermitian dynamics of the spin-1 XY model\,\cite{PhysRevLett.123.147201}, one of the paradigmatic QMBS models, and show that adding certain non-Hermitian terms significantly stabilize the non-thermal dynamics.

\subsection{Algebraic structure of QMBS}
Following Ref.\,\cite{PhysRevLett.119.030601}, we introduce the SM construction of QMBS Hamiltonians: let $\hat{P}_{[i]}$ be a local projector acting on the site(s) around $i$. We consider the following form of Hamiltonian $\hat{H}_{\mr{SM}}$,
\begin{equation}\label{eq:SM form}
    \hat{H}_{\mr{SM}}=\sum_i\hat{P}_{[i]}\hat{h}_{[i]}\hat{P}_{[i]}+\hat{H}',
\end{equation}
where $\hat{h}_{[i]}$ is an arbitrary Hermitian operator and $\hat{H}'$ should satisfy the following condition,
\begin{equation}
    [\hat{P}_{[i]},\hat{H}']=0,\,\,\forall i.
\end{equation}
Let $\mc{S}\coloneqq\bigcap_i\ker\hat{P}_{[i]}$ be the common kernel of these projectors, i.e., $\hat{P}_{[i]}\mc{S}=0$ for $\forall i$, which we call the scar subspace. $\mc{S}$ is by construction invariant under $\hat{H}_{\mr{SM}}$ (i.e., $\hat{H}_{\mr{SM}}\mc{S}\subset\mc{S}$), since 
\begin{equation}
    \hat{H}_{\mr{SM}}\mc{S}=\hat{H}'\mc{S},\,\,\hat{P}_{[i]}\hat{H}'\mc{S}=\hat{H}'\hat{P}_{[i]}\mc{S}=0\Rightarrow\hat{H}'\mc{S}\subset\mc{S}.
\end{equation}
Therefore, if $\mc{S}$ is not empty, at least one exact eigenstate of $\hat{H}_{\mr{SM}}$ lies entirely within $\mc{S}$. Such eigenstates are referred to as scar states. Ref.\,\cite{PhysRevLett.119.030601} further shows rigorously that the scar states of this Hamiltonian violate the ETH. 

The local projector $\hat{P}_{[i]}$ is typically classified based on the size of its support: in the simpler class, $\hat{P}_{[i]}$ is a one-body operator, i.e., $\hat{P}_{[i]}\equiv\hat{P}^{\mr{SM}}_i$. For example, the spin-1 XY model\,\cite{PhysRevLett.123.147201} and the model discussed in Ref.\,\cite{PhysRevE.96.022153} contain such projectors. The more common class consists of two-body projectors $\hat{P}_{[i]}\equiv\hat{P}^{\mr{SM}}_{i,i+1}$. Many toy Hamiltonians of QMBS contain such projectors, and they usually project out the anti-symmetric part of the wavefunction, upon a suitable local invertible transformation. 

The term $\hat{H}'$ is typically the Zeeman term, i.e., a translationally invariant single-body (non-interacting) term. Indeed, for $\hat{P}_{[i]}$ annihilating the anti-symmetric part of the wavefunction, one can show that the only admissible form of $\hat{H}'$ is the Zeeman term, as long as the locality of the Hamiltonian is imposed~\cite{KOPhD}. 

One can easily generalize this construction by relaxing the condition on the first term in Eq.\,\eqref{eq:SM form}. For example, we can consider
\begin{equation}\label{eq:SM general}
    \hat{H}'_{\mr{SM}}=\sum_i\hat{h}_{[i]}\hat{P}_{[i]}+\hat{H}',
\end{equation}
where $\hat{h}_{[i]}$ can be long-ranged and non-Hermitian. Trivially $\mc{S}$ is an invariant subspace for this model as well. Though this extension seems hardly relevant for physically plausible models, i.e., local Hermitian Hamiltonians, Eq.\,\eqref{eq:SM general} appears in several well-known (Hermitian) models, which we will introduce in the following.

\subsection{Non-Herminicity in QMBS}
While many QMBS models can be easily cast into the SM form [Eq.\,\eqref{eq:SM form}], identifying this form ($\hat{H}=\sum_i\hat{P}_{[i]}\hat{h}_{[i]}\hat{P}_{[i]}+\hat{H}'$) is not obvious for certain models. Such models have been regarded as ``beyond SM'', and it was speculated that more elaborate model construction is necessary. The most well-known examples include the PXP model, the effective spin-1/2 chain of the Rydberg experiment, 
\begin{equation}
    \hat{H}_{\mr{PXP}}=\sum_{i=1}^L\hat{P}_{i-1}\hat{X}_i\hat{P}_{i+1},
\end{equation}
where $\hat{P}_i\coloneqq\dyad*{\downarrow}$ is the projector onto the down-spin (which is not the projector $\hat{P}_i^{\mr{SM}}$ in Eq.\,\eqref{eq:SM form}) and $\hat{X}_i\coloneqq\dyad*{\uparrow}{\downarrow}+\dyad*{\downarrow}{\uparrow}$ is the $x$-component of the Pauli matrices. However, Refs.\,\cite{PhysRevA.107.023318,PhysRevB.108.054412} have shown that the SM form emerges once we add the following non-Hermitian perturbation,
\begin{equation}
    \hat{H}'_{\mr{PXP}}\coloneqq\hat{H}_{\mr{PXP}}+\delta\hat{H}_{\mr{NH}}\coloneqq\hat{H}_{\mr{PXP}}+\frac{1}{4}\sum_{i=1}^L\hat{P}_{i-1}\hat{\sigma}_i^+\hat{P}_{i+1}\left(\hat{P}_{i-2}+\hat{P}_{i+2}\right),
\end{equation}
where $\hat{\sigma}_i^+\coloneqq\dyad*{\uparrow}{\downarrow}$ is the raising operator. The exact scar states of this deformed PXP model are a good approximation of those of the original PXP model (see also Appendix\,\ref{appendix:pxp} for a brief account). The SM form of the deformed PXP model is hidden in the enlarged Hilbert space, 
\begin{equation}\label{eq:H'PXP decomp}
    \hat{H}'_{\mr{PXP}}\hat{P}_{\mr{Ryd}}=\hat{P}_{\mr{Ryd}}\left(\sum_{i=1}^L\hat{h}_{[i]}\widetilde{P}_{i,i+1}+\hat{H}_Z\right)\eqqcolon\hat{P}_{\mr{Ryd}}\widetilde{H}'_{\mr{PXP}},
\end{equation}
where $\hat{P}_{\mr{Ryd}}\coloneqq\prod_{i=1}^L(1-\dyad*{\uparrow\uparrow}_{i,i+1})$ is the Rydberg projector, $\hat{h}_{[i]}$ is a certain non-Hermitian term, $\widetilde{P}_{i,i+1}$ is a two-body projector, and $\hat{H}_Z$ is a non-Hermitian Zeeman term. The SM structure is easily identified in $\widetilde{H}'_{\mr{PXP}}$, and one can therefore obtain the corresponding exact scar states, which we denote as $\ket*{\widetilde{S}_n}$. The scar state of the deformed PXP model is then obtained by applying the Rydberg projector onto these states, i.e., $\ket{S_n}=\hat{P}_{\mr{Ryd}}\ket*{\widetilde{S}_n}$. 

Another example is the spin-1/2 model consisting of Dzyaloshinski-Moriya (DM) interaction, 
\begin{equation}
    \hat{H}_{\mr{DM}}\coloneqq D\sum_{i=1}^L\left(\hat{X}_i\hat{Z}_{i+1}-\hat{Z}_i\hat{X}_{i+1}\right),
\end{equation}
where $\hat{Z}\coloneqq\dyad*{\uparrow}-\dyad*{\downarrow}$ is the $z$-component of the Pauli matrices. This model has the simple ferromagnetic states $\ket{S_n}=(\sum_i\hat{\sigma}_i^+)^n\ket{\downarrow\cdots\downarrow}$ as exact zero-energy scar states. $\hat{H}_{\mr{DM}}$ can be cast into the generalized SM form Eq.\,\eqref{eq:SM general}, where $\hat{h}'_{[i]}$ becomes a long-ranged non-Hermitian term with its maximum support $L$~\cite{PhysRevB.108.054412}. 

The examples above showcase emergent non-Herminicity even in Hermitian models at the mathematical level, implying a certain connection between QMBS and non-Hermitian physics. 

\subsection{Simulation of the non-Hermitian spin-1 XY model}
Finally, we simulate the non-Hermitian dynamics of the paradigmatic spin-1 XY model with some perturbations that disrupt its SM structure. We note that the idea of stabilizing QMBS by non-Hermitian terms was first proposed in Ref.\,\cite{PhysRevLett.133.216601}, which studies the many-body revivals in the PXP model numerically. Here, we consider a theoretically simpler setting, but take a different type of non-Hermitian term which makes the spectrum scatter in the complex plane. In Sec.\,\ref{sec:numerics on XY}, we will argue that the stabilized non-thermal dynamics from our non-Hermitian perturbation persist \textit{indefinitely}.

The clean spin-1 XY model on the one-dimensional chain is defined as follows~\cite{PhysRevLett.123.147201},
\begin{equation}
    \hat{H}_{\mr{XY}}=J\sum_{i=1}^{L-1}\left(\hat{S}_i^x\hat{S}_{i+1}^y+\hat{S}_i^y\hat{S}_{i+1}^y\right)+\sum_{i=1}^L\left(h\hat{S}_i^z+D(\hat{S}_i^z)^2\right),
\end{equation}
where $\hat{S}_i^\alpha$ is the spin-1 operator. For (numerical) simplicity we assume that the system has the open boundary and the size of the chain $L$ is even. We define the basis that diagonalizes the $\hat{S}^z$ operator as $\{\ket{\pm},\ket{0}\}$, where they satisfy $\hat{S}^z\ket{\pm}=\pm\ket{\pm}$ and $\hat{S}^z\ket{0}=0$. The scar states of this Hamiltonian is written as,
\begin{equation}
    \ket{S_n}\coloneqq\frac{1}{N_n}\left(\sum_{i=1}^L(-1)^i(\hat{S}_i^+)^2\right)^n\bigotimes_{j=1}^L\ket{-}_j,
\end{equation}
where $N_n$ is the normalization factor and $\hat{S}^\pm\coloneqq\hat{S}^x\pm i\hat{S}^y$ is the spin-raising/lowering operator. This model can be cast into the SM form, with the following projectors $\hat{P}_i^{\mr{XY}}$ and $\hat{P}^{\mr{XY}}_{i,i+1}$, and the Hermitian term $\hat{h}_{i,i+1}^{\mr{XY}}$,
\begin{equation}\label{eq:xy annihilator}
\begin{split}
    \hat{P}^{\mr{XY}}_i&\coloneqq\dyad*{0}_i\\
    \hat{P}^{\mr{XY}}_{i,i+1}&\coloneqq\frac{1}{2}\big(\ket*{+,-}+\ket*{-,+}\big)\big(\bra*{+,-}+\bra*{-,+}\big)_{i,i+1}\\
    \hat{h}_{i,i+1}^{\mr{XY}}&\coloneqq\big(\dyad*{+,0}{0,+}+\dyad*{0,+}{+,0}+\dyad*{-,0}{0,-}+\dyad*{0,-}{-,0}\big)_{i,i+1}\\
    &+\ket*{0,0}\big(\bra{+,-}+\bra*{-,+}\big)_{i,i+1}+\big(\ket*{+,-}+\ket*{-,+}\big)\bra*{0,0}_{i,i+1}.
\end{split}
\end{equation}
As the initial state of the quench dynamics, we consider the following anti-ferromagnetic product state,
\begin{equation}
    \ket{\psi_{\mr{AF}}}\coloneqq\prod_{l=1}^{L/2}\frac{1}{2}\left(\ket{+}+\ket{-}\right)_{2l-1}\otimes\left(\ket{+}-\ket{-}\right)_{2l}.
\end{equation}
One can easily verify that $\ket*{\psi_{\mr{AF}}}$ is a superposition of the scar states $\{\ket*{S_n}\}_n$, and thus the dynamics ensuing from this state never thermalizes. To measure the non-thermal nature of the system's dynamics more quantitatively, we define the ``order parameter'' $\hat{O}$,
\begin{equation}
    \hat{O}\coloneqq\frac{1}{L}\sum_{i=1}^L\left(1-\hat{P}^{\mr{XY}}_i\right).
\end{equation}
One can show rigorously that the expectation value of $\hat{O}$
with respect to the Gibbs ensemble is strictly smaller than unity ($\mr{tr}\hat{\rho}_{\mr{Gibbs}}\hat{O}<1$), while that of the dynamics ensuing from $\ket{\psi_{\mr{AF}}}$ remains unity at an arbitrary time ($\expval*{\hat{O}(t)}{\psi_{\mr{AF}}}=1$), demonstrating ergodicity breaking.

\begin{figure}
    \begin{minipage}{0.4\textwidth}
        \centering
        \includegraphics[width=\textwidth]{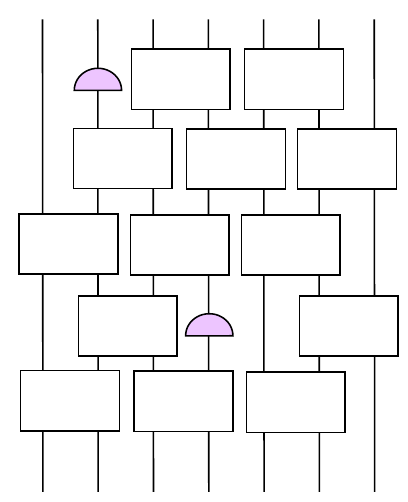}
    \end{minipage}
    \begin{minipage}{0.58\textwidth}
        \centering
    \includegraphics[width=\textwidth]{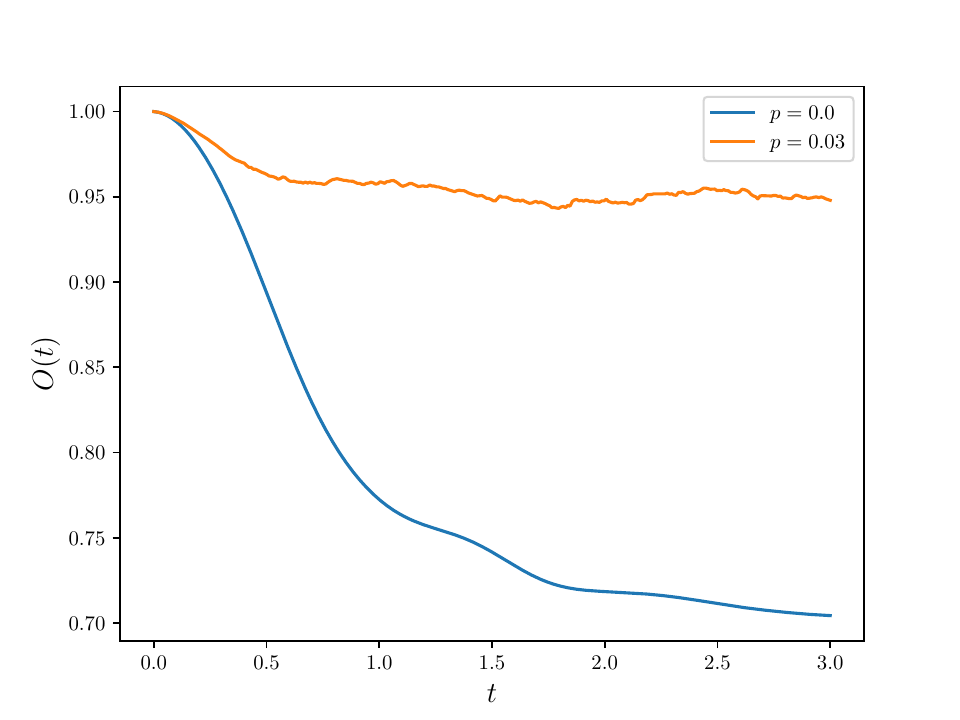}
    \end{minipage}
    \caption{Simulation of the non-unitary dynamics of the spin-1 XY model ensuing from the $\ket*{\psi_{\mr{AF}}}$ state. (Left) A graphical summary of the simulation: each update of the wavefunction by $e^{-it\hat{h}_{i,i+1}}$ (in Eq.\,\eqref{eq:h_ii+1}) is replaced with the projector $1-\hat{P}^{\mr{XY}}_i$ or $1-\hat{P}^{\mr{XY}}_{i+1}$ with the probability $p$. (Right) The time-evolution of the order parameter $O(t)$ in Eq.\,\eqref{eq:orderparam}. For the simulation we set the parameters as $J=h=D=V=1.0$, and $L=20$. The dynamics is approximated by TEBD with the MPS ansatz of the bond dimension $\chi=200$ (we have simulated the system varying $\chi$ from $100$ to $200$, and the dynamics seems to converge for the time window we consider). When the stochastic measurement has been implemented with the non-zero probability, we run $30$ independent simulations and take the sample average.}
    \label{fig:pxp simulation}
\end{figure}

Let us add the following simple perturbation $\hat{V}$ that disrupts the SM structure,
\begin{equation}
    \hat{V}\coloneqq V\sum_{i=1}^L\hat{S}_i^x.
\end{equation}
Consequently, $\ket{S_n}$ ceases to be an exact eigenstate. Indeed, $\hat{V}$ does not commute with either $\hat{P}^{\mr{XY}}_i$ or $\hat{P}^{\mr{XY}}_{i,i+1}$, invalidating the assumption of the SM construction. We numerically observe that the state $\ket*{\psi_{\mr{AF}}}$ does not exhibit non-thermal dynamics anymore, and instead eventually becomes thermalized (see the right panel of Fig.\,\ref{fig:pxp simulation}). To stabilize the non-thermal dynamics ensuing from $\ket*{\psi_{\mr{AF}}}$, we apply the following projector stochastically,
\begin{equation}
    \ket*{\psi(t)}\mapsto\left(1-\hat{P}^{\mr{XY}}_i\right)\ket*{\psi(t)},
\end{equation}
with the probability $g$ in the unit time, i.e., $P_{\mr{m}}(t,t+dt)=gdt$, where $P_{\mr{m}}(t,t')$ is the probability of applying the projector in the time interval $[t,t']$. This insertion does not affect the scar states by construction, while it projects out non-scar configurations, increasing the relative weight of the scar states in the wavefunction. This process is equivalent to adding the following non-Hermitian perturbation,
\begin{equation}
    \hat{H}_{\mr{NH}}=-ig\sum_{i=1}^L\hat{P}^{\mr{XY}}_i.
\end{equation}

We numerically study the quantum trajectory of the dynamics under the projective ``measurement'' $1-\hat{P}^{\mr{XY}}_i$ from the initial state $\ket{\psi_{\mr{AF}}}$,
\begin{equation}
    \ket*{\psi(t)}=e^{-i\left(\hat{H}_{\mr{XY}}+\hat{V}\right)t_M}\prod_{a=0}^{M-1}\left(\left(1-\hat{P}^{\mr{XY}}_{i_a}\right)e^{-i\left(\hat{H}_{\mr{XY}}+\hat{V}\right)t_a}\right)\ket{\psi_{\mr{AF}}},
\end{equation}
where the projector is randomly inserted at a randomly-chosen site $i_a$ after randomly-selected elapsed time $t_a$, according to the probability distribution, i.e., $t_0+\cdots+t_M=t$. The dynamics is simulated by time-evolving block decimation (TEBD)~\cite{AJDaley_2004} (see the left panel of Fig.\,\ref{fig:pxp simulation} for a graphical summary of the setup). In the right panel of Fig.\,\ref{fig:pxp simulation}, we plot the expectation value of the ``order parameter'' for the non-thermalization, 
\begin{equation}\label{eq:orderparam}
    O(t)\coloneqq\frac{1}{L}\sum_{i=1}^L\frac{\expval{\hat{O}_i}{\psi(t)}}{\innerproduct{\psi(t)}}.
\end{equation} 
To simulate the dynamics of the system by TEBD, we replace the Trotterized two-site local unitary gate $\hat{U}_{i,i+1}=e^{-it\hat{h}_{i,i+1}}$ with the projector $1-\hat{P}^{\mr{XY}}_i$ or $1-\hat{P}^{\mr{XY}}_{i+1}$ with the probability $p$. Here the two-site local operator $\hat{h}_{i,i+1}$ is defined as,
\begin{equation}\label{eq:h_ii+1}
    \hat{h}_{i,i+1}\coloneqq J\left(\hat{S}_i^x\hat{S}_{i+1}^y+\hat{S}_i^y\hat{S}_{i+1}^y\right)+h\left(\hat{S}_i^z+\hat{S}_{i+1}^z\right)+D\left((\hat{S}_i^z)^2+(\hat{S}_{i+1}^z)^2\right)+V\left(\hat{S}_i^x+\hat{S}_{i+1}^x\right).
\end{equation}
Without insertion of the projector, i.e., $p=0$, the order parameter $O(t)$ rapidly decays to the stationary value, indicating fast thermalization. This typical behavior, however, is drastically modified by the measurement: even with sporadic insertions of the projector with $p=0.03$, the order parameter stays close to the unity, and ergodicity remains broken for a much longer time.

The numerics above demonstrates the stabilized non-thermal behavior of the scar states within the matrix product state (MPS) approximation. Since the volume law entanglement cannot be efficiently expressed by the MPS, the late time behavior of the system is not fully reliable. Nevertheless, this observation motivates us to speculate that with more frequent ``measurements'', or a strong non-Hermitian coupling equivalently, the non-thermal dynamics would persist even indefinitely, making the scar states true non-equilibrium steady state. We will show in the following that this scenario indeed holds for certain models.

\section{Random Quantum Circuit Model}
\label{sec:rqc}
We consider a one dimensional random quantum circuit (RQC) model that supports a scar subspace arising from the SM structure. The circuit consists of $L$ qudits, with its local Hilbert space $\mf{h}=\mb{C}^q$. We denote the local basis by $\ket{a}$ ($a=1\sim q$). We also define the ``local scar subspace'' as $\mf{s}=\mb{C}^{|\mf{s}|}$ ($|\mf{s}|<q$) with the basis $\{\ket{a}; a=1\sim|\mf{s}|\}$. We define the (whole) scar subspace as the tensor product of the local scar subspaces, i.e., $\mc{S}=\bigotimes_{i=1}^L\mf{s}$, whereas the total Hilbert space is $\bigotimes_{i=1}^L\mf{h}$. We assume that $L$ is even for simplicity.

Before proceeding, let us consider the dynamics under the following simple QMBS Hamiltonian that makes $\mc{S}$ as the scar subspace in the sense of the SM construction,
\begin{equation}
    \hat{H}_0=\sum_{i=1}^L\hat{\Pi}_{i,i+1}\hat{h}_{i,i+1}\hat{\Pi}_{i,i+1}+\sum_{i=1}^L\hat{h}'_i,
\end{equation}
where 
\begin{equation}
    \hat{\Pi}_{i,i+1}\coloneqq\left(1-\sum_{a\in\mf{s}}\dyad{a}_i\right)\left(1-\sum_{b\in\mf{s}}\dyad{b}_{i+1}\right),
\end{equation}
is a two-body projector annihilating the scar subspace, $\hat{h}_{i,i+1}$ is an arbitrary local Hermitian term acting non-trivially on the sites $i$ and $i+1$, and $\hat{h}'_i$ is a one-body operator that commutes with $\hat{\Pi}_{j,j+1}$ for $\forall j$. From Eq.\,\eqref{eq:SM form} it is clear that $\mc{S}$ is the scar subspace of this model. We further add the following perturbation $\hat{V}$ and the non-Hermitian term $\hat{H}_{\mr{NH}}$,
\begin{equation}
    \begin{split}
        \hat{V}&=\sum_{i=1}^L\hat{V}_{i,i+1}\\
        \hat{H}_{\mr{NH}}&=-ig\sum_{i=1}\left(1-\hat{P}_{i,i+1}\right),
    \end{split}
\end{equation}
where $\hat{V}_{i,i+1}$ is a local operator acting non-trivially on the sites $i$ and $i+1$, which does not preserve the SM structure, and $\hat{P}_{i,i+1}$ is the following two-body projector onto the scar subspace,
\begin{equation}\label{eq:P}
    \hat{P}_{i,i+1}\coloneqq\left(\sum_{a\in\mf{s}}\dyad{a}_i\right)\left(\sum_{b\in\mf{s}}\dyad{b}_{i+1}\right).
\end{equation}
These terms model the system in Sec.\,\ref{sec:motivation} such that $\hat{V}$ disrupts the SM structure and $\hat{H}_{\mr{NH}}$ restores it.
In a suitable rotating frame, the Zeeman term $\sum_{i=1}^L\hat{h}'_i$ can be eliminated, at the cost of introducing a time-dependent Hamiltonian,
\begin{equation}\label{eq:H(t) RQC}
    \begin{split}
        \hat{H}(t)&=\hat{H}_0(t)+\hat{V}(t)+\hat{H}_{\mr{NH}}\\
        \hat{H}_0(t)&=\sum_{i=1}^L\hat{\Pi}_{i,i+1}\hat{h}_{i,i+1}(t)\hat{\Pi}_{i,i+1}\\
        \hat{V}(t)&=\sum_{i=1}^L\hat{V}_{i,i+1}(t),
    \end{split}
\end{equation}
where $\hat{h}_{i,i+1}(t)$ and $\hat{V}_{i,i+1}(t)$ are the corresponding local time-dependent operators. Roughly speaking, we will model the dynamics under $\hat{H}(t)$ in Eq.\,\eqref{eq:H(t) RQC} by a simple circuit model in the following, replacing $e^{-i\Delta t\hat{H}(t)}$ with a product of local random unitary operators.

Our circuit system evolves under the stochastic combination of three operations: two-local random unitary gates $\hat{U}$ that preserves the local scar subspace, two-local fully Haar random unitary $\hat{V}$, and the measurement $\hat{P}$. The unitary $\hat{U}$ models the dynamics under $\hat{H}_0(t)$ in Eq.\,\eqref{eq:H(t) RQC}, 
\begin{equation}
    \hat{U}_{i,i+1}\coloneqq\sum_{a,b\in\mf{s}}\dyad{a,b}_{i,i+1}+\sum_{(a,b)\not\in\mf{s}^{\otimes2}}\sum_{(c,d)\not\in\mf{s}^{\otimes2}}U_{ab,cd}\dyad{a,b}{c,d}_{i,i+1},
\end{equation}
where $\mf{s}$ is identified with the set $\mf{s}=\{1\sim |\mf{s}|\}$. The above form ensures that the scar subspace is invariant under this operation ($\hat{U}\mf{s}^{\otimes2}\subset\mf{s}^{\otimes 2}$). The matrix elements $U_{ab,cd}$ in the second term are drawn from $U(q^2-|\mf{s}|^2)$ Haar random unitary, satisfying the following property\footnote{The Haar random unitary means much more than Eq.\,\eqref{eq:Haar random}, but we only use Eq.\,\eqref{eq:Haar random} in this paper.}, 
\begin{equation}\label{eq:Haar random}
    \int\mu(dU)U_{ab,cd}U^*_{a'b',c'd'}=\frac{1}{q^2-s^2}\delta_{aa'}\delta_{bb'}\delta_{cc'}\delta_{dd'},\,\,(a,b),(a',b'),(c,d),(c',d')\not\in\mf{s}^{\otimes2}.
\end{equation}
The gate $\hat{V}$ disrupts the scar subspace and mixes the local scar subspace $\mf{s}$ and its ``thermal'' complement $\mf{h}\setminus\mf{s}$, which allows one to interpret it as a ``scar-breaking perturbation'' $\hat{V}(t)$ in Eq.\,\eqref{eq:H(t) RQC}, 
\begin{equation}
    \hat{V}_{i,i+1}\coloneqq\sum_{a,b,c,d\in\mf{q}}V_{ab,cd}\dyad{ab}{cd}_{i,i+1},
\end{equation}
where $\mf{q}$ is identified with the set $\mf{q}=\{1\sim q\}$. Just like the unperturbed gate $\hat{U}$, we assume that $\hat{V}$ satisfies the following statistical property,
\begin{equation}
    \int\mu(dV)V_{ab,cd}V^*_{a'b',c'd'}=\frac{1}{q^2}\delta_{aa'}\delta_{bb'}\delta_{cc'}\delta_{dd'}.
\end{equation}
Finally, the operator $\hat{P}$ (Eq.\,\eqref{eq:P}) locally projects the state onto the local scar subspace.

The system is updated by the quantum circuit consisting of $\hat{U}$, $\hat{V}$, and $\hat{P}$. At each step, one of them is chosen to update the state according to the probability distribution: we will apply the projector $\hat{P}_{i,i+1}$ on each pair of sites with the probability $c$, the unperturbed gate $\hat{U}$ with the probability $a$, and the perturbation $\hat{V}$ with the probability $b=1-a-c$. Here $b$ can be understood as the perturbation strength. The average time evolution of the system $\hat{\rho}$ is expressed as the following (trace non-preserving) quantum channel $\mc{E}_{i,i+1}$,
\begin{equation}\label{eq:channel}\begin{split}
    \mc{E}_{i,i+1}(\hat{\rho})&=a\hat{U}_{i,i+1}\hat{\rho}\hat{U}^\dag_{i,i+1}+b\hat{V}_{i,i+1}\hat{\rho}\hat{V}^\dag_{i,i+1}+c\hat{P}_{i,i+1}\hat{\rho}\hat{P}_{i,i+1}.
    \end{split}
\end{equation}

We take the average over a corresponding unitary group for each unitary gate, and the resulting channel is denoted as $\mc{L}_{i,i+1}$, 
\begin{equation}
    \mc{L}_{i,i+1}(\hat{\rho})=\int\mu(d\hat{U}_{i,i+1})a\hat{U}_{i,i+1}\hat{\rho}\hat{U}^\dag_{i,i+1}+\int\mu(d\hat{V}_{i,i+1})b\hat{V}_{i,i+1}\hat{\rho}\hat{V}^\dag_{i,i+1}+c\hat{P}_{i,i+1}\hat{\rho}\hat{P}_{i,i+1}.
\end{equation}
The many-body time-evolution is generated by products of these channels, i.e., $\mc{L}_{\mr{odd}}\coloneqq\mc{L}_{1,2}\circ\cdots\circ\mc{L}_{N-1,N}$ and $\mc{L}_{\mr{even}}\coloneqq\mc{L}_{2,3}\circ\cdots\circ\mc{L}_{N,1}$. The non-equilibrium steady state is therefore defined as the eigenstate of the operator $\mc{L}\coloneqq\mc{L}_{\mr{even}}\circ\mc{L}_{\mr{odd}}$ corresponding to the eigenvalue with the largest real part.

\subsection{Mapping to the spin-1/2 model}
\subsubsection{Formulation}
The model Eq.\,\eqref{eq:channel} can be mapped to the spin-1/2 model for certain initial states. To see this, let us consider the maximally mixed local state within the full Hilbert space and within the local scar subspace,
\begin{equation}
    \hat{q}\coloneqq\frac{1}{q}\sum_{a=1}^q\dyad*{a},\,\,\hat{s}\coloneqq\frac{1}{|\mf{s}|}\sum_{a\in\mf{s}}\dyad*{a}.
\end{equation}
The dynamics is closed within the subspace of the density matrices spanned by $\hat{q}$ and $\hat{s}$. Namely, if the initial state density matrix can be expanded by these density matrices, the density matrix at an arbitrary time can also be described as a linear combination of them. Indeed the time evolution is written as,
    \begin{equation}\label{eq:action L}\begin{split}
        \mc{L}_{i,i+1}(\hat{s}_i\hat{s}_{i+1})&=(a+c)\hat{s}_i\hat{s}_{i+1}+b\hat{q}_i\hat{q}_{i+1}\\
        \mc{L}_{i,i+1}(\hat{s}_i\hat{q}_{i+1})&=(a+c)r\hat{s}_i\hat{s}_{i+1}+\left(a\left(1-r\right)+b\right)\hat{q}_i\hat{q}_{i+1}\\
        \mc{L}_{i,i+1}(\hat{q}_i\hat{s}_{i+1})&=(a+c)r\hat{s}_i\hat{s}_{i+1}+\left(a\left(1-r\right)+b\right)\hat{q}_i\hat{q}_{i+1}\\
        \mc{L}_{i,i+1}(\hat{q}_i\hat{q}_{i+1})&=cr^2\hat{s}_i\hat{s}_{i+1}+(a+b)\hat{q}_i\hat{q}_{i+1},
        \end{split}
    \end{equation}
with $r\coloneqq|\mf{s}|/q$.

Equation\,\eqref{eq:action L} further allows us to map the model into an effective spin-1/2 model, by identifying $\hat{q}$ and $\hat{s}$ with the up-spin and the down-spin, respectively,
\begin{equation}\label{eq:mapping state}
    \hat{q}\mapsto\ket{\uparrow}_{\mr{eff}},\,\,\hat{s}\mapsto\ket{\downarrow}_{\mr{eff}}.
\end{equation}
With this identification, the operator $\mc{L}_{i,i+1}$ is mapped to $\hat{L}_{i,i+1}\in\mc{B}(\mb{C}^{2\otimes2})$, 
\begin{equation}\label{eq:mapping channel}\begin{split}
    \hat{L}_{i,i+1}&=\begin{pmatrix}
        \ket{\uparrow\uparrow}_{\mr{eff}}&\ket{\downarrow\downarrow}_{\mr{eff}}&\ket{\uparrow\downarrow}_{\mr{eff}}&\ket{\downarrow\uparrow}_{\mr{eff}}
    \end{pmatrix}_{i,i+1}
    \begin{pmatrix}
        \msf{A} & \msf{B} \\ \msf{O} & \msf{O}
    \end{pmatrix}
    \begin{pmatrix}
        \bra{\uparrow\uparrow}_{\mr{eff}} \\ \bra{\downarrow\downarrow}_{\mr{eff}} \\ \bra{\uparrow\downarrow}_{\mr{eff}} \\ \bra{\downarrow\uparrow}_{\mr{eff}}
    \end{pmatrix}_{i,i+1}\\
    \msf{A}&=\begin{pmatrix}
        a+b&b\\
        cr^2&a+c
    \end{pmatrix},\,\,
    \msf{B}=\begin{pmatrix}
      a(1-r)+b&a(1-r)+b\\
      (a+c)r&(a+c)r
    \end{pmatrix}.
    \end{split}
\end{equation}

The non-equilibrium steady state is the eigenstate(s) of $\hat{L}\coloneqq\prod_l\hat{L}_{2l-1,2l}\prod_l\hat{L}_{2l,2l+1}$ corresponding to the eigenvalue with the largest real part. 

\subsubsection{Exactly solvable limits}\label{subsubsec:limiting cases}
The steady state of $\hat{L}$ can be obtained for the limiting cases of $b=0$ and $c=0$, corresponding to the scarred steady state and the thermal steady state, respectively.

\noindent\textbf{The case $b=0$}: The local operator $\hat{L}_{i,i+1}$ has at most two eigenstates with the eigenvalue $1$: if $c=0$, the states $\ket{\downarrow\downarrow}_{\mr{eff}}$ and $\ket{\uparrow\uparrow}_{\mr{eff}}$ possess the eigenvalue with the largest real part,
and if $c\not=0$ only $\ket{\downarrow\downarrow}_{\mr{eff}}$ has the eigenvalue with the largest real part. 
Since the absolute value of any eigenvalue of $\hat{L}_{i,i+1}$ is bounded by $1$, and the same holds for $\hat{L}$, we obtain 
\begin{equation}\label{eq:exact sol scar}
    \begin{dcases}
        \ket{\downarrow\cdots\downarrow}_{\mr{eff}},\,\,\ket{\uparrow\cdots\uparrow}_{\mr{eff}}\,\,&(c=0)\\
        \ket{\downarrow\cdots\downarrow}_{\mr{eff}}\,\,&(c\not=0)
    \end{dcases},
\end{equation}
as the steady state(s). Note that $\ket{\downarrow\cdots\downarrow}_{\mr{eff}}$ corresponds to the density matrix $\hat{s}\otimes\cdots\otimes\hat{s}$, i.e., the state that lies within the scar subspace, while $\ket{\uparrow\cdots\uparrow}_{\mr{eff}}$ corresponds to the maximally mixed thermal state. Intuitively, Eq.\,\eqref{eq:exact sol scar} indicates that, in the absence of the perturbation, even an infinitesimal measurement ($c>0$) drives the system into the scar subspace.

\noindent\textbf{The case $c=0$}: A similar discussion above shows that the steady state is 
\begin{equation}
    \begin{dcases}
        \ket{\downarrow\cdots\downarrow}_{\mr{eff}},\,\,\ket{\uparrow\cdots\uparrow}_{\mr{eff}}\,\,&(b=0)\\
        \ket{\uparrow\cdots\uparrow}_{\mr{eff}}\,\,&(b\not=0)
    \end{dcases}.
\end{equation}

We note that in our setting the density matrix which can be expressed as a linear combination of $\hat{q}$ and $\hat{s}$ should always exhibit volume-law entanglement entropy. Indeed, the entanglement entropy $S_l(t)$ of $l$ consecutive sites (say $1\sim l$) for an arbitrary time $t$ is bounded as 
\begin{equation}\label{eq:entanglement bound}
    |\mf{s}|l\leq S_l(t),
\end{equation}
for $\forall t$.

\subsection{Mapping to a classical statistical model}
\begin{figure}
    \begin{minipage}{0.49\textwidth}
        \centering
        \includegraphics[width=\linewidth]{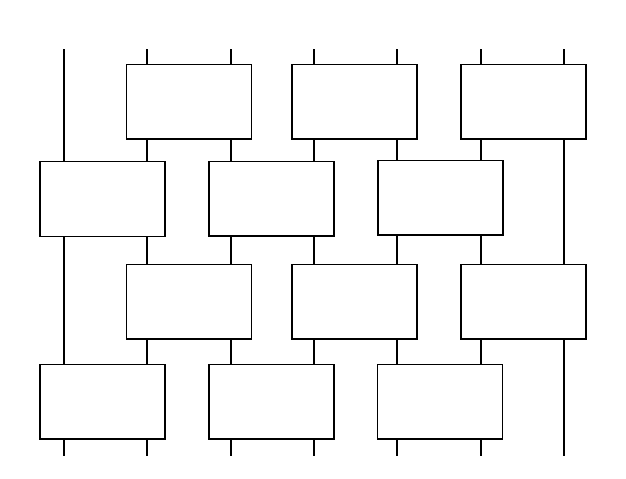}
    \end{minipage}
    \begin{minipage}{0.49\textwidth}
    \centering
    \includegraphics[width=\linewidth]{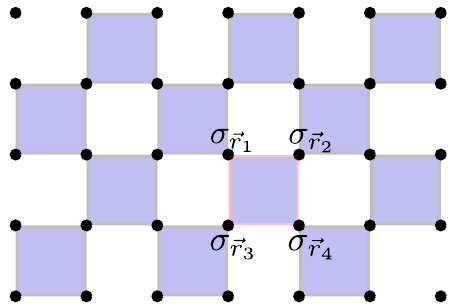}
    \end{minipage}
    \caption{The circuit architecture (left panel) is mapped to the two dimensional ``spacetime'' lattice. In the two dimensional classical model, the dynamical variable is the Ising spin $\sigma_{\vec{r}}=\pm1$, and the interaction or the Boltzmann weight $W$ is defined on the filled plaquette.}
    \label{fig:lattice}
\end{figure}
For general values of $b$ and $c$, the analytic solution is not available. To obtain the steady state approximately, we map the problem into the two dimensional classical model. 
Given that the initial density matrix is a linear combination of $\hat{q}$ and $\hat{s}$, the density matrix (or the wavefunction upon the mapping Eq.\,\eqref{eq:mapping state} and Eq.\,\eqref{eq:mapping channel}) after a time $T$ is written as, 
\begin{equation}\label{eq:time evolution}
    \mc{L}^T(\hat{\rho}_{\mr{ini}})\leftrightarrow\hat{L}^T\ket{\rho_{\mr{ini}}}_{\mr{eff}}
\end{equation}
For simplicity we assume that $T$ is an integer multiple of the system size $L$. This assumption will be used in Sec.\,\ref{subsec:approximation} to compute the steady state properties of the system. Equation\,\eqref{eq:time evolution} motivates us to define the following partition function,
\begin{equation}\label{eq:partition function}
    Z\coloneqq\mr{tr}_{\mr{eff}}\hat{L}^T=\sum_{\{\sigma_{\Vec{r}}\}_{\Vec{r}}}\prod_{\square}W[\{\sigma_{\vec{r}}\}_{\vec{r}\in\square}],
\end{equation}
where $\Vec{r}=(i,t)$ is the ``spacetime'' coordinate consisting of the spatial site index $i$ and the time slice $t$. The trace in Eq.\,\eqref{eq:partition function} is taken over the effective spin-1/2 degrees of freedom ($\ket{\uparrow}_\mr{eff}$ and $\ket{\downarrow}_\mr{eff}$), which is equivalent to imposing periodic boundary conditions (PBC) in both the spatial and temporal directions.

The ``Boltzmann weight'' $W[\{\sigma_{\vec{r}}\}]$ is equivalent to that of the two dimensional classical model on the square lattice, which is defined on plaquette (see the right panel of Fig.\,\ref{fig:lattice}),
\begin{equation}\label{eq:Ising}\begin{split}
    W[\{\sigma_{\vec{r}}\}_{\vec{r}\in\square}]
    &\equiv W\Big(\plaquetteconfigwide{\sigma_{\vec{r}_1}}{\sigma_{\vec{r}_2}}{\sigma_{\vec{r}_3}}{\sigma_{\vec{r}_4}}\Big)\coloneqq _{\mr{eff}}\!\!\!\mel{\sigma_{\vec{r}_1},\sigma_{\vec{r}_2}}{\hat{L}_{i,i+1}}{\sigma_{\vec{r}_3},\sigma_{\vec{r}_4}}_{\mr{eff}},
    \end{split}
\end{equation}
where $\sigma_{\vec{r}}=\uparrow,\downarrow$ is the Ising variable. Similarly, Eq.\,\eqref{eq:partition function} implies that the system at the time $t$ is described as
\begin{equation}
    \ket{\rho(t)}_{\mr{eff}}=\sum_{\{\sigma_{\vec{r}}\}}\prod_{\square}W[\{\sigma_{\vec{r}}\}]\ket*{\{\sigma_{(i,t)}\}_i}_{\mr{eff}}.
\end{equation}
Analogous to equilibrium statistical physics, certain properties of the steady state can be inferred from the ``free energy'' $F\coloneqq-T^{-1}\ln Z$, evaluated in the limit of large $T$. In the following, we provide an approximation of this quantity.

\subsection{Transfer matrix method}
\begin{figure}
    \centering
    \includegraphics[width=0.6\linewidth]{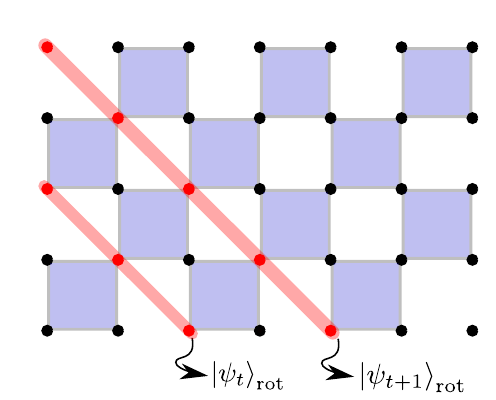}
    \caption{The sites on the red line are identified as the time slice with the corresponding quantum state $\ket{\psi_t}_{\mr{rot}}$. The time-evolution is generated by the transfer matrix [Eq.\,\eqref{eq:U}].}
    \label{fig:lattice mps}
\end{figure}
\begin{figure}
    \begin{minipage}{0.59\linewidth}
    \centering
    \includegraphics[width=\textwidth]{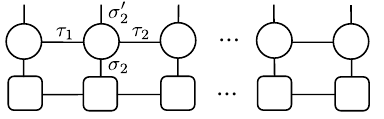}
    \end{minipage}
    \begin{minipage}{0.4\linewidth}
        \centering
        \includegraphics[width=\textwidth]{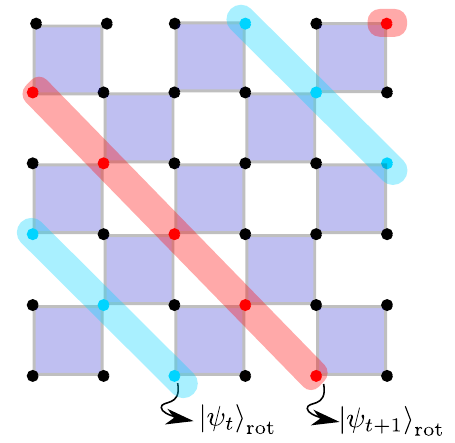}
    \end{minipage}
    \caption{Left panel: the dynamics can be expressed as the MPO acting on the MPS. Note that the lines connecting the edges due to the periodic boundary are omitted in this diagram. Right panel: the time-evolution of the state $\ket{\psi_{t+1}}_{\mr{rot}}=\hat{U}\ket{\psi_t}_{\mr{rot}}$ in the original spacetime lattice (before rotating the quantization axis) for $L=T=6$. Particularly, $\hat{U}$ is well-defined when $T$ is an integer multiple of $L$.}
    \label{fig:mps mpo}
\end{figure}

To approximate the partition function $Z$, we regard the sites in the diagonal line as the time slice with the corresponding quantum state $\ket{\psi(t)}_{\mr{rot}}$ (see the red sites in Fig.\,\ref{fig:lattice mps}), i.e., we rotate the quantization axis by $\pi/4$. The adjacent diagonal line right to the time slice $t$ is then identified with the next discrete time slice $t+1$ with the corresponding state $\ket*{\psi(t+1)}_{\mr{rot}}$. The transfer matrix $\hat{U}$ generating the time-evolution is defined as (see also the left panel of Fig.\,\ref{fig:mps mpo}),
\begin{equation}\label{eq:U}
    _{\mr{rot}}\!\!\mel{\sigma'_1\cdots\sigma'_L}{\hat{U}}{\sigma_1\cdots\sigma_L}_{\mr{rot}}=\sum_{\{\tau_i\}_i}W\Big(\plaquetteconfig{\tau_1}{\sigma'_1}{\sigma_1}{\tau_2}\Big)\cdots W\Big(\plaquetteconfig{\tau_L}{\sigma'_L}{\sigma_L}{\tau_1}\Big).
\end{equation}
We note that, as $T$ is assumed to be an integer multiple of the system size $L$, the time-evolution $\hat{U}$ is well-defined (see the right panel of Fig.\,\ref{fig:mps mpo} for the case of $L=T=6$).

Equation\,\eqref{eq:U} indicates that the partition function is written as successive applications of the transfer matrix $\hat{U}$ onto the quantum state $\ket{\psi_t}_{\mr{rot}}$ with the periodic boundary condition in time and space. This implies that, by writing $\ket*{\psi_t}_{\mr{rot}}$ as the matrix product state (MPS), the ``time-evolution'' is expressed as the simple diagram using the matrix product operator (MPO) depicted in Fig.\,\ref{fig:mps mpo}. Thus, the free energy per time is equivalent to the eigenvalue with the largest real part $\lambda$ of $\hat{U}$ (we assume that $\hat{U}$ is invertible). The corresponding equation is diagrammatically written as 
\begin{equation}\label{eq:eigenequation}\begin{split}
    \hat{U}\ket{\Phi}_{\mr{rot}}&\equiv\includegraphics[valign=c,width=0.35\linewidth]{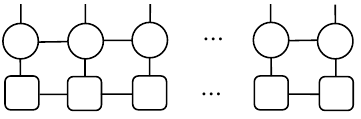}=\lambda\includegraphics[valign=c,width=0.35\linewidth]{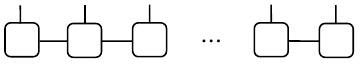}\\
    &=\lambda\ket{\Phi}_{\mr{rot}}.
    \end{split}
\end{equation}

To find the maximum eigenvalue of $\hat{U}$ approximately, we cast Eq.\,\eqref{eq:eigenequation} into the variational problem, 
\begin{equation}\label{eq:variational principle}\begin{split}
    |\lambda|^2&=\max_{\ket{\Phi}_{\mr{rot}}}\frac{_{\mr{rot}}\!\expval*{\hat{U}^\dag\hat{U}}{\Phi}_{\mr{rot}}}{_{\mr{rot}}\!\innerproduct*{\Phi}{\Phi}_{\mr{rot}}}=\max_{\ket*{\Phi}_{\mr{rot}}}\frac{\includegraphics[valign=c,width=0.213\linewidth]{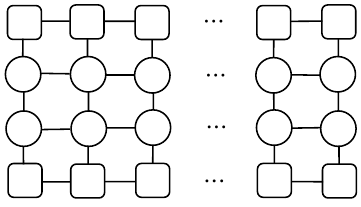}}{\includegraphics[valign=c,width=0.213\linewidth]{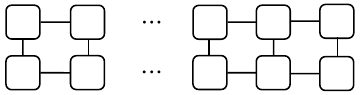}}\\
    &=\max_{\ket{\Phi}_{\mr{rot}};_{\mr{rot}}\!\innerproduct*{\Phi}_{\mr{rot}}=1}\includegraphics[valign=c,width=0.22\linewidth]{figures/mps_mpo_mps_pbc.pdf}.
    \end{split}
\end{equation}
Furthermore, we regard the RHS of Eq.\,\eqref{eq:variational principle} as the partition function of a one-dimensional classical model, i.e., the multiplication of the transfer matrices $\hat{K}$. 
\begin{equation}\label{eq:kernel}
    \includegraphics[valign=c,width=0.25\linewidth]{figures/mps_mpo_mps_pbc.pdf}=\mr{tr}_{\mr{rot}}\hat{K}^{\blue{T}}=\mr{tr}_{\mr{rot}}\left[\includegraphics[valign=c,width=0.05\linewidth]{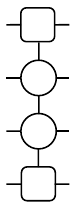}\right]^{\blue{T}},
\end{equation}
where the trace is taken over the Hilbert space of spin-1/2 degrees of freedom ($\ket{\uparrow}_{\mr{rot}}$ and $\ket{\downarrow}_{\mr{rot}}$) in the time slice after the rotation of the quantization axis.

\subsection{Variational optimization of Eq.\,\eqref{eq:variational principle}}
We first numerically optimize Eq.\,\eqref{eq:variational principle} to identify the steady state, based on the mean-field approximation and the density matrix renormalization group (DMRG), both of which provide consistent results (Fig.\,\ref{fig:mf diagram}). They demonstrate the existence of two distinct phases for the steady state, corresponding to the scarred state and the thermal state. In Sec.\,\ref{sec:RG}, we instead investigate the partition function itself by the real space renormalization group (RG) approach. 

\subsubsection{Mean-field approximation}\label{subsec:approximation}
We first try to find the optimal $\lambda$ in Eq.\,\eqref{eq:variational principle} within a rather simple class of MPS, 
\begin{equation}\begin{split}
    \ket{\psi}_{\mr{rot}}&=\sum_{\{\sigma_i\}_i}\mr{tr}\left(\msf{A}^{\sigma_1}\cdots\msf{A}^{\sigma_L}\right)\ket*{\sigma_1\cdots\sigma_L}_{\mr{rot}}\\
    \msf{A}^\uparrow&=e^{i\phi/2}\cos\theta,\,\,\msf{A}^\downarrow=e^{-i\phi/2}\sin\theta,
    \end{split}
\end{equation}
which corresponds to a mean-field approximation, where $_{\mr{rot}}\!\expval*{\hat{U}^\dag\hat{U}}{\Phi}_{\mr{rot}}$ is optimized with respect to the product state $\ket*{\Phi}_{\mr{rot}}$. With this ansatz, the largest eigenvalue $\tau(\theta,\phi)$ of $\hat{K}$ is,
\begin{equation}
    \tau(\theta,\phi)=\frac{A+D}{2}+\frac{1}{2}\sqrt{(A-D)^2+4BC},
\end{equation}
where 
\begin{equation}\begin{split}
    A&=\left|e^{-i\phi/2}\cos\theta W\Big(\plaquetteconfigwide{\uparrow}{\uparrow}{\uparrow}{\uparrow}\Big)+e^{i\phi/2}\sin\theta W\Big(\plaquetteconfigwide{\uparrow}{\uparrow}{\downarrow}{\uparrow}\Big)\right|^2\\
    B&=\left|e^{-i\phi/2}\cos\theta W\Big(\plaquetteconfigwide{\uparrow}{\uparrow}{\uparrow}{\downarrow}\Big)+e^{i\phi/2}\sin\theta W\Big(\plaquetteconfigwide{\uparrow}{\uparrow}{\downarrow}{\downarrow}\Big)\right|^2\\
    C&=\left|e^{-i\phi/2}\cos\theta W\Big(\plaquetteconfigwide{\downarrow}{\downarrow}{\uparrow}{\uparrow}\Big)+e^{i\phi/2}\sin\theta W\Big(\plaquetteconfigwide{\downarrow}{\downarrow}{\downarrow}{\downarrow}\Big)\right|^2\\
    D&=\left|e^{-i\phi/2}\sin\theta W\Big(\plaquetteconfigwide{\downarrow}{\downarrow}{\uparrow}{\uparrow}\Big)+e^{i\phi/2}\sin\theta W\Big(\plaquetteconfigwide{\downarrow}{\downarrow}{\downarrow}{\downarrow}\Big)\right|^2,
    \end{split}
\end{equation}
with the corresponding eigenvector,
\begin{equation}\label{eq:eigvec mf}
    \ket{\psi(\theta,\phi)}_{\mr{rot}}\propto B\ket{\uparrow}_{\mr{rot}}+\frac{1}{2}\left(\sqrt{(A-D)^2+4BC}-(A-D)\right)\ket{\downarrow}_{\mr{rot}}.
\end{equation}

\begin{figure}[h]    
    \begin{minipage}[b]{0.49\linewidth}
        \centering
    \includegraphics[width=\linewidth]{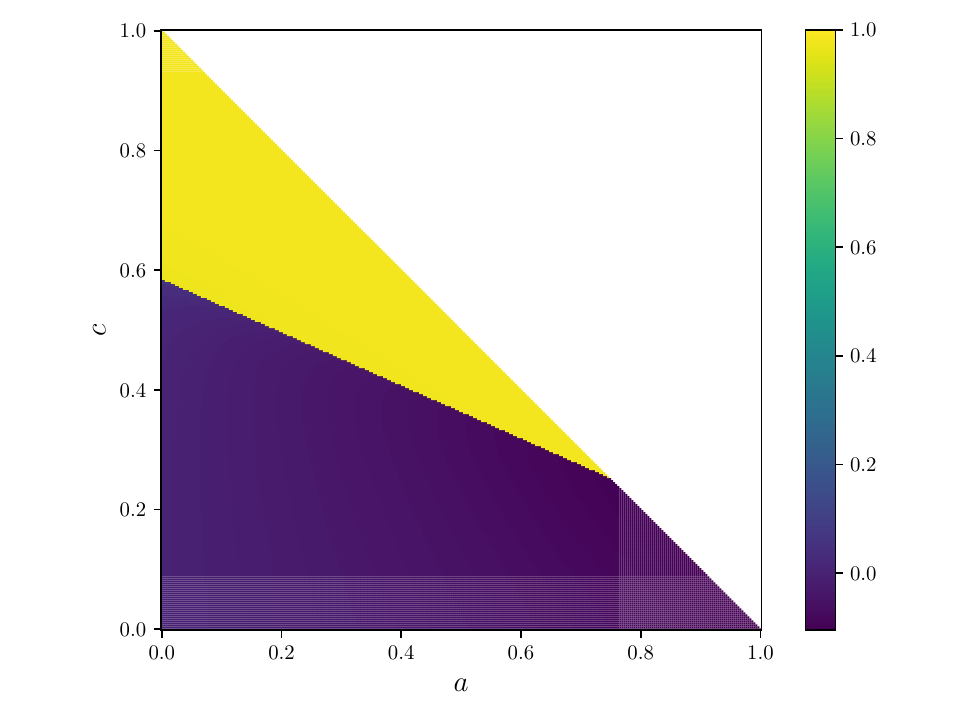}
    \end{minipage}
\hfill
    \begin{minipage}[b]{0.49\linewidth}
        \centering
        \includegraphics[width=\linewidth]{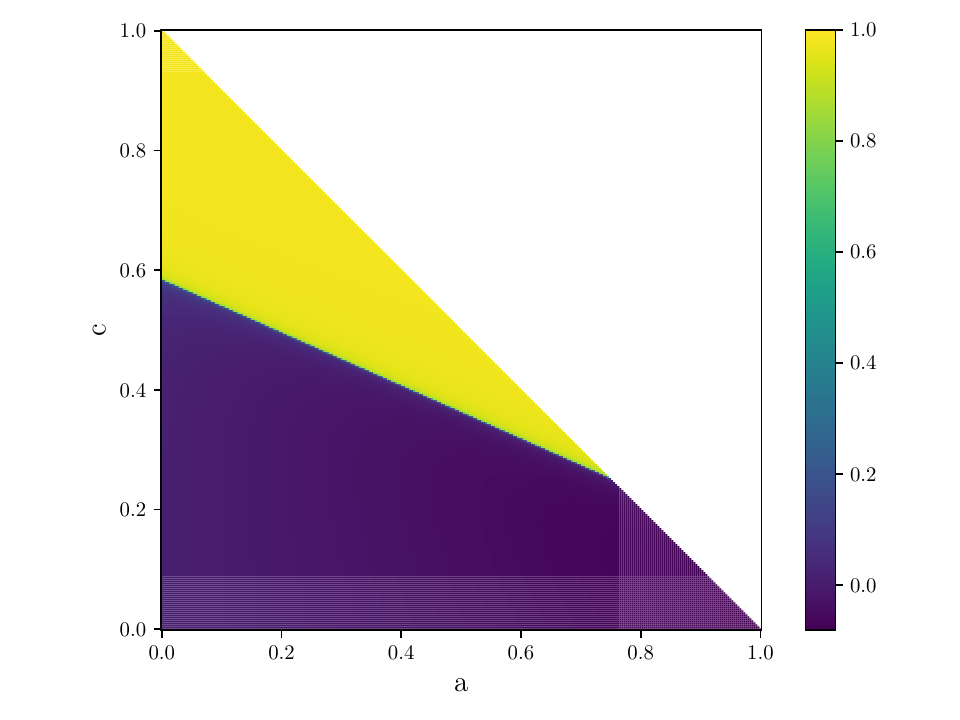}
    \end{minipage}

    \caption{The expectation value of the magnetization $M(\theta,\phi)$ with respect to the optimal state within the mean-field calculation (the left panel), and within DMRG (the right panel). Note that $a$ and $c$ should satisfy the conditions $0\leq a,c$ and $0\leq a+c\leq1$.}
    \label{fig:mf diagram}
\end{figure}

Since the up-spin and the down-spin correspond to the thermal and the scarred states, respectively, the expectation value of the ``magnetization'',
\begin{equation}
    \hat{M}_{\mr{eff}}\coloneqq-_{\mr{eff}}\!\dyad*{\uparrow}_{\mr{eff}}+_{\mr{eff}}\!\dyad{\downarrow}_{\mr{eff}},
\end{equation}
serves as a measure of the weight of the steady state in the scar subspace: $\expval*{\hat{M}_{\mr{eff}}}=1$ indicates that the steady state is completely scarred, while $\expval*{\hat{M}_{\mr{eff}}}=-1$ implies the thermal steady state. Note that the magnetization should be evaluated \textit{before} rotating the quantization axis, i.e., $_{\mr{eff}}\!\expval*{\hat{M}_{\mr{eff}}}{\psi}_{\mr{eff}}$. Nevertheless, we compute, 
\begin{equation}
    M(\theta,\phi)\coloneqq _{\mr{rot}}\!\!\!\expval*{\hat{M}_{\mr{rot}}}{\psi(\theta,\phi)}_{\mr{rot}},
\end{equation}
where $\hat{M}_{\mr{rot}}$ is the corresponding magnetization operator for the rotated quantization axis (see the left panel of Fig.\,\ref{fig:mf diagram}). The results reveal the existence of two distinct phases: For a sufficiently large $c$, the magnetization $M_{\mr{rot}}(\theta,\phi)$ approaches unity, indicating that the steady state lies almost entirely within the scar subspace; we refer to this regime as the \textit{QMBS phase}. In contrast, for small $c$ (or large $b$), the magnetization is close to zero. Although this nearly zero magnetization differs from the naive expectation $M_{\mr{rot}}(\theta,\phi)\approx-1$ for the thermal state, we nonetheless identify this regime with \textit{thermal phase}. A more detailed discussion will be presented in Sec.\,\ref{sec:RG}.



\subsubsection{Density matrix renormalization group}
\label{sec:DMRG}
We directly optimize the state $\ket{\Phi}_{\mr{rot}}$ in Eq.\,\eqref{eq:variational principle} by DMRG, where the operator $\hat{U}^\dag\hat{U}$ is expressed as the MPO and the MPS ansatz $\ket*{\Phi}_{\mr{rot}}$ is optimized locally in order to maximize the expectation value $_{\mr{rot}}\!\expval*{\hat{U}^\dag\hat{U}}{\Phi}_{\mr{rot}}$~\cite{PhysRevLett.69.2863}. We vary the bond dimension of the MPS ansatz $\ket*{\Phi}_{\mr{rot}}$ from $\chi=100$ to $\chi=200$, which exhibits clearly visible convergence. The expectation value of the magnetization $_{\mr{rot}}\!\!\expval*{\hat{M}_{\mr{rot}}}{\Phi}_{\mr{rot}}$ (the right panel of Fig.\,\ref{fig:mf diagram}) even quantitatively agrees with the mean-field approximation, confirming the existence of the QMBS and the thermal phases. 

Furthermore, we find that both the mean-field calculation and DMRG underestimate the QMBS phase: the exact solution Eq.\,\eqref{eq:exact sol scar} indicates that the whole diagonal line except for the point $(a,c)=(1,0)$ in Fig.\,\ref{fig:mf diagram} should be in the QMBS phase, meaning that even an infinitesimal $c$ can break ergodicity, whereas in our approximations this phase appears only when the ``measurement'' probability is large enough. This behavior is in common with the other approach based on real space renormalization group described below, and we attribute it to the rotated quantization axis (Fig.\,\ref{fig:lattice mps}), which alters the quantum channel from $\hat{L}$ to $\hat{U}$. Indeed, one can find that the exact steady state $\ket{\uparrow\cdots\uparrow}_{\mr{eff}}$ or $\ket{\downarrow\cdots\downarrow}_{\mr{eff}}$ (\textit{before} rotating the quantization axis) in the solvable limiting cases in Sec.\,\ref{subsubsec:limiting cases} does not solve the equation Eq.\,\eqref{eq:variational principle}, i.e., neither $\ket{\uparrow\cdots\uparrow}_{\mr{rot}}$ nor $\ket{\downarrow\cdots\downarrow}_{\mr{rot}}$ is an exact eigenstate of $\hat{U}$. In Sec.\,\ref{sec:eft}, we address this limitation by considering the complementary approach: instead of rotating the quantization axis, we modify the local Hilbert space of the qudit such that the model becomes analytically tractable.

We emphasize that this transition is distinct from the measurement induced phase transition (MIPT) which separates the thermal state and the low entanglement state in a similar manner. In our model,  both phases have the volume law entanglement due to the bound Eq.\,\eqref{eq:entanglement bound}. 
\subsection{Real space renormalization group}
\label{sec:RG}
\begin{figure}
    \centering
    \begin{minipage}{0.49\textwidth}
        \includegraphics[width=\linewidth]{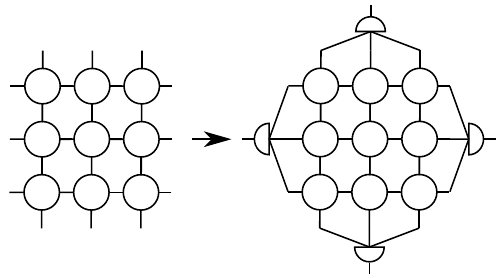}
    \end{minipage}
    \begin{minipage}{0.49\textwidth}
        \includegraphics[width=\linewidth]{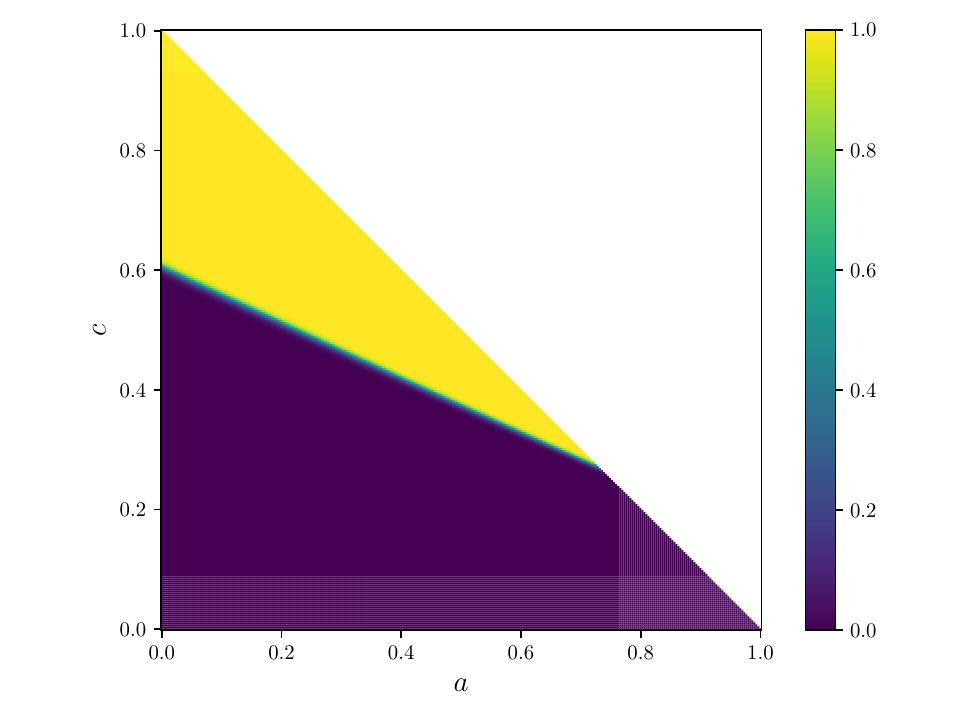}
    \end{minipage}
    \caption{Left panel: a tensor network representation of the RG procedure, where the circles indicate the Boltzmann weights in the MPO representation. Right panel: phase diagram obtained by the real space RG analysis. We plot the value of the component $\sigma_{\vec{r}_1}=\sigma_{\vec{r}_2}=\sigma_{\vec{r}_3}=\sigma_{\vec{r}_4}=\downarrow$ (\eqref{eq:qmbs fixed pt}) in the Boltzmann weight, which becomes unity in the QMBS phase. The yellow region indicates that after the RG transformations, the Boltzmann weight flows to the QMBS fixed point (Eq.\,\eqref{eq:qmbs fixed pt}). The purple region indicates that the Boltzmann weight flows to the thermal phase Eq.\,\eqref{eq:thermal fixed pt}.}
    \label{fig:RG}
\end{figure}

The other approach is based on the real space RG~\cite{Cardy_1996}. The procedure is the same as the block spin transformation for the two dimensional Ising model, which is summarized in the tensor network notation in Fig.\,\ref{fig:RG}. We define the block spin variable by the majority rule of $3\times3$ original spins, 
\begin{equation}\label{eq:majority rule}
    \widetilde{\sigma}_{\vec{r}}=\begin{dcases}
        \uparrow\,\,&\sigma_{\vec{r}-\vec{e}}+\sigma_{\vec{r}}+\sigma_{\vec{r}+\vec{e}}>0\\
        \downarrow\,\,&\sigma_{\vec{r}-\vec{e}}+\sigma_{\vec{r}}+\sigma_{\vec{r}+\vec{e}}<0
    \end{dcases}\equiv\includegraphics[valign=c,width=0.15\linewidth]{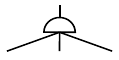},
\end{equation}
where $\vec{r}$ is the space-time coordinate in the original setting and $\vec{e}\coloneqq(+1,-1)^T$. In the condition of Eq.\,\eqref{eq:majority rule}, we identify $\sigma_{\vec{r}}=\uparrow$ and $\sigma_{\vec{r}}=\downarrow$ with $+1$ and $-1$, respectively. By contracting the Boltzmann weights in the $3\times3$ network by the majority rule Eq.\,\eqref{eq:majority rule} attached to its edges, we define the new Boltzmann weight $\widetilde{W}$,
\begin{equation}\label{eq:RG procedure}\begin{split}
    &W\Big(\plaquetteconfig{\sigma_1}{\sigma_2}{\sigma_3}{\sigma_4}\Big)\mapsto\widetilde{W}\Big(\plaquetteconfig{\widetilde{\sigma}_1}{\widetilde{\sigma}_2}{\widetilde{\sigma}_3}{\widetilde{\sigma}_4}\Big)
    \end{split}
\end{equation}
The last step in the RG procedure is rescaling: one of the non-zero Boltzmann weights is set to be unity and the other weights are redefined as a ratio between these weights, i.e., $\widetilde{W}\mapsto\mc{R}[W]\coloneqq\widetilde{W}/W_c$ with $W_c$ the reference value.

Despite too many couplings ($2^4-1=15$ components of the Boltzmann weight, where $-1$ comes from rescaling) that seems to render the RG analysis formidable, by studying the RG flow numerically for various initial points $(a,b,c)$, we empirically identify only two relevant directions in which the Boltzmann weight can flow within the possible values of the parameters $a,b$ and $c$, each corresponding to the QMBS phase and the thermal phase.

\subsubsection{QMBS fixed points}
We find the following simple stable fixed point:
\begin{equation}\label{eq:qmbs fixed pt}
    W\Big(\plaquetteconfigwide{\downarrow}{\downarrow}{\downarrow}{\downarrow}\Big)=1,
\end{equation}
and any other element is zero. This fixed point trivially corresponds to the QMBS state. Its stability is easily verified analytically by the linearized RG equation: denoting the fixed point as $W^*$, the RG equation of the Boltzmann weight slightly away from $W^*$ is 
\begin{equation*}
    \mc{R}[W^*+\delta W]=W^*+\order{\delta W^2},
\end{equation*}
where $\delta W$ is a small deviation. This confirms the stability of this fixed point.

\subsubsection{Thermal phase}
The other direction corresponds to thermalization: under the RG, the Boltzmann weight flows to one with the following non-zero component,  
\begin{equation}\label{eq:thermal fixed pt}
    W\Big(\plaquetteconfigwide{\uparrow}{\uparrow}{\downarrow}{\downarrow}\Big)=1.
\end{equation}
Any other component becomes exponentially small, and eventually negligible as the block spin transformations are applied sufficiently many times, i.e., a runaway RG flow that signals a discontinuous phase transition. We further numerically find that there are three more ``slow'' directions: while the system asymptotically flow to the Boltzmann weight Eq.\,\eqref{eq:thermal fixed pt} in the limit of infinitely many RG steps, after a finite but large RG transformations, the following components remain non-zero,
\begin{equation}
    W\Big(\plaquetteconfigwide{\uparrow}{\uparrow}{\downarrow}{\uparrow}\Big)\eqqcolon x,\,\,W\Big(\plaquetteconfigwide{\uparrow}{\uparrow}{\uparrow}{\downarrow}\Big),\,\,W\Big(\plaquetteconfigwide{\uparrow}{\uparrow}{\uparrow}{\uparrow}\Big)\eqqcolon y,
\end{equation}
while the other components are negligible.

These four nonzero components in the Boltzmann weight corresponds to the thermal phase, as the transfer matrix forces all the spins to be up,
\begin{equation}
   \hat{U}=\sum_{\{\sigma_i\}_i}\sum_{\{\tau_i\}_i}W\Big(\plaquetteconfigwide{\tau_1}{\uparrow}{\sigma_1}{\tau_2}\Big)\cdots W\Big(\plaquetteconfigwide{\tau_L}{\uparrow}{\sigma_L}{\tau_1}\Big)\,\, _{\mr{rot}}\!\dyad{\uparrow\cdots\uparrow}{\sigma_1\cdots\sigma_L}_{\mr{rot}}.
\end{equation}
This form also provides a plausible scenario for the nearly zero magnetization observed in the mean-field and DMRG results: the variational problem [Eq.\,\eqref{eq:variational principle}] is exactly solvable for above $\hat{U}$, and the expectation value of the magnetization is written as, 
\begin{equation}
    \expval*{\hat{M}_{\mr{rot}}}=-\frac{x^2-y^2}{x^2+y^2},
\end{equation}
which becomes nearly zero when $x$ and $y$ are of the same order. 

In the right panel of Fig.\,\ref{fig:RG}, we plot the value of the component of the Boltzmann weight that is nonzero in the QMBS phase ($\sigma_1=\sigma_2=\sigma_3=\sigma_4=\downarrow$ in Eq.\,\eqref{eq:Ising}) after five RG transformations. Numerically, five RG steps are 
already sufficient to observe to which direction the Boltzmann weight flows. The diagram obtained by the real space RG is even quantitatively consistent with the mean-field and the DMRG results, further confirming the existence of the first order phase transition and the QMBS phase.


\section{Effective field theory}
\label{sec:eft}
We have discussed the phase transition between the QMBS phase and the thermal phase by mapping the RQC dynamics into the two dimensional classical model. In this section we obtain some analytical understanding of the nature of this transition by constructing an effective field theory. To this end, we regard the qudit ($\mf{h}=\mb{C}^q$) in the RQC model as a fundamental representation of $SU(q)$. To construct the semi-classical ``free energy functional'', we instead identify the qudit with a different irrep of $SU(q)$, and consider a similar $SU(q)$ spin model to Eq.\,\eqref{eq:H(t) RQC} or Eq.\,\eqref{eq:channel}. In the large-$q$ limit, this model turns out to be equivalent to an $SU(1,1)$ spin chain with a large Bragmann index, an analog of spin quantum number, which permits the semi-classical analysis.  

We consider the following time-dependent Hamiltonian,
\begin{equation}\label{eq:H}
    \begin{split}
        \hat{H}(t)&=\hat{H}_0+\hat{H}_{\mr{NH}}+\hat{V}_{\mr{pert}}\\
        \hat{H}_0(t)&=\sum_{i=1}^L\sum_{(\mu,\rho)\not\in\mf{s}^2}\sum_{(\nu,\sigma)\not\in\mf{s}^2}J_i^{\mu\nu\rho\sigma}(t)\hat{S}_{i,\mu\nu}\hat{S}_{i+1,\rho\sigma}\\
        \hat{H}_{\mr{NH}}&=-ig\sum_{i=1}^L\sum_{\mu\not\in\mf{s}}\hat{S}_{i,\mu\mu}\\
        \hat{V}_{\mr{pert}}(t)&=\sum_{i=1}^L\sum_{\mu,\rho\not\in\mf{s}}\sum_{\nu,\sigma\in\mf{s}}V_i^{\mu\nu\rho\sigma}(t)\hat{S}_{i,\mu\nu}\hat{S}_{i+1,\rho\sigma}+h.c.
    \end{split}
\end{equation}
$\hat{S}_{i,\mu\nu}$ is the $SU(q)$ spin operator that satisfies the $\mf{su}(q)$ algebra at each site. The structure of the model Eq.\,\eqref{eq:H} is similar to that of Eq.\,\eqref{eq:H(t) RQC} (see the discussion below). In Eq.\,\eqref{eq:H}, we assume that the local Hilbert space is the totally symmetric combination of $\kappa q$ fundamentals, i.e., the corresponding Young tableau consists of $\kappa q$ boxes in a single row. Here, $\kappa$ is a fixed $O(1)$ quantity ($0<\kappa$), which makes the following analysis self-consistent. This irrep is conveniently represented by the Schwinger boson $\hat{a}_{i\mu}$,
\begin{equation}\label{eq:Schwinger}
    \hat{S}_{i,\mu\nu}=\hat{a}^\dag_{i\mu}\hat{a}_{i\nu},\,\,\sum_{\mu=1}^q\hat{a}_{i\mu}^\dag\hat{a}_{i\mu}=\kappa q,
\end{equation}
where $\hat{a}_{i\mu}(\hat{a}^\dag_{i\mu})$ is the boson annihilation (creation) operator. The local scar subspace is then defined as,
\begin{equation}
    \mf{s}\coloneqq\mr{span}\left\{\prod_{\mu\in\mf{s}}\frac{\left(\hat{a}_\mu^\dag\right)^{n_\mu}}{\sqrt{n_\mu!}}\ket*{\mr{vac}};\,\,\sum_{\mu\in\mf{s}}n_\mu=\kappa q\right\},
\end{equation}
where $\ket{\mr{vac}}$ is the vacuum state in the boson Fock space. The scar subspace $\mc{S}\coloneqq\bigotimes_{i=1}^L\mf{s}$ is the tensor product of the local scar subspace. One can confirm that $\mc{S}$ is invariant under the action of $\hat{H}_0$. Furthermore, the non-Hermitian term is written as $\hat{H}_{\mr{NH}}=-ig\sum_i\sum_{\mu\not\in\mf{s}}\hat{a}_{i\mu}^\dag\hat{a}_{i\mu}$ which forces non-scar components to decay.

The coupling constants $J_{ij}^{\mu\nu\rho\sigma}$ and $V_{ij}^{\mu\nu\rho\sigma}$ are drawn from the Gaussian,
\begin{equation}\label{eq:disorder}\begin{split}
    \overline{J_i^{\mu\nu\rho\sigma}(t)}&=0,\,\,\overline{J_i^{\mu\nu\rho\sigma}(t)(J_{i'}^{\mu'\nu'\rho'\sigma'}(t'))^*}=\frac{2U^2}{q^3}\delta(t-t')\delta_{ii'}\delta_{\mu\mu'}\delta_{\nu\nu'}\delta_{\rho\rho'}\delta_{\sigma\sigma'}\\
    \overline{V_i^{\mu\nu\rho\sigma}(t)}&=0,\,\,\overline{V_i^{\mu\nu\rho\sigma}(t)(V_{i'}^{\mu'\nu'\rho'\sigma'}(t'))^*}=\frac{2V^2}{q^3}\delta(t-t')\delta_{ii'}\delta_{\mu\mu'}\delta_{\nu\nu'}\delta_{\rho\rho'}\delta_{\sigma\sigma'}.
    \end{split}
\end{equation}

\subsection{The effective time evolution}
Following the RQC setup, we consider the time-evolution of the density matrix initialized with the maximally mixed state $\hat{\rho}\propto\hat{1}$, assuming that the steady state is independent of the initial state. The equation of motion is, 
\begin{equation}
    i\dv{\hat{\rho}(t)}{t}=\llbracket\hat{H}(t),\hat{\rho}(t)\rrbracket,
\end{equation}
where $\llbracket\hat{A},\hat{B}\rrbracket$ is the generalized commutator for non-Hermitian dynamics,
\begin{equation}
    \llbracket\hat{A},\hat{B}\rrbracket\coloneqq\hat{A}\hat{B}-\hat{B}\hat{A}^\dag.
\end{equation}
One can follow the derivation of the master equation to obtain the following effective equation of motion after the disorder average by Eq.\,\eqref{eq:disorder}, 
\begin{equation}\label{eq:master eq}
    \dv{\overline{\hat{\rho}(t)}}{t}=-i\Big\llbracket\hat{H}_{\mr{NH}},\overline{\hat{\rho}(t)}\Big\rrbracket+\mc{D}[\overline{\hat{\rho}(t)}],
\end{equation}
where the bar indicates the disorder average and $\mc{D}[\hat{\rho}]$ is a super-operator that describes the dissipation (see Appendix\,\ref{appsec:master equation} for the derivation),
\begin{equation}\label{eq:dissipator}
    \begin{split}
        \mc{D}\big[\hat{\rho}\big]&\coloneqq\frac{2J^2}{q^3}\sum_{i=1}^L\sum_{(\mu,\rho)\not\in\mf{s}^2}\sum_{(\nu,\sigma)\not\in\mf{s}^2}\left[\hat{S}_{i,\mu\nu}\hat{S}_{i+1,\rho\sigma}\hat{\rho}\hat{S}_{i,\nu\mu}\hat{S}_{i+1,\sigma\rho}-\frac{1}{2}\left\{\hat{S}_{i,\mu\nu}\hat{S}_{i,\nu\mu}\hat{S}_{i+1,\rho\sigma}\hat{S}_{i+1,\sigma\rho},\hat{\rho}\right\}\right]\\
        &+\frac{2V^2}{q^3}\sum_{i=1}^L\sum_{\mu\rho\not\in\mf{s}}\sum_{\nu\sigma\in\mf{s}}\left[\hat{S}_{i,\mu\nu}\hat{S}_{i+1,\rho\sigma}\hat{\rho}\hat{S}_{i,\nu\mu}\hat{S}_{i+1,\sigma\rho}-\frac{1}{2}\left\{\hat{S}_{i,\mu\nu}\hat{S}_{i,\nu\mu}\hat{S}_{i+1,\rho\sigma}\hat{S}_{i+1,\sigma\rho},\hat{\rho}\right\}\right]\\
        &+\frac{2V^2}{q^3}\sum_{i=1}^L\sum_{\mu\rho\not\in\mf{s}}\sum_{\nu\sigma\in\mf{s}}\left[\hat{S}_{i,\nu\mu}\hat{S}_{i+1,\sigma\rho}\hat{\rho}\hat{S}_{i,\mu\nu}\hat{S}_{i+1,\rho\sigma}-\frac{1}{2}\left\{\hat{S}_{i,\nu\mu}\hat{S}_{i,\mu\nu}\hat{S}_{i+1,\sigma\rho}\hat{S}_{i+1,\rho\sigma},\hat{\rho}\right\}\right].
    \end{split}
\end{equation}

We examine the steady state solution of Eq.\,\eqref{eq:master eq}. To this end, we map the density matrix to the ket vector by the Choi–Jamio\l kowski isomorphism $\overline{\hat{\rho}(t)}\mapsto\ket*{\rho(t)}$. By this isomorphism, the $SU(q)$ spin operators are mapped to,
\begin{equation}
    \hat{S}_{i,\mu\nu}\hat{\rho}\mapsto\left(\hat{S}_{i,\mu\nu}\otimes\hat{1}\right)\ket*{\rho},\,\,\hat{\rho}\hat{S}_{i,\mu\nu}\mapsto\left(\hat{1}\otimes\hat{S}_{i,\nu\mu}\right)\ket*{\rho},
\end{equation} 
where $\hat{1}$ is the identity operator acting on either the ``ket'' space or the ``bra'' space. Following the Keldysh formalism, we denote $\hat{S}_{i,\mu\nu}\otimes\hat{1}\eqqcolon\hat{S}^+_{i,\mu\nu}$ and $\hat{1}\otimes\hat{S}_{i,\mu\nu}\eqqcolon\hat{S}^-_{i,\mu\nu}$. The equation of motion Eq.\,\eqref{eq:master eq} is then mapped to the Schr\"odinger equation,
\begin{equation}\label{eq:Schrodinger eq}
    \dv{t}\ket*{\rho(t)}=-g\sum_{i=1}^L\sum_{\mu\not\in\mf{s}}\left(\hat{S}^+_{i,\mu\mu}+\hat{S}^-_{i,\mu\mu}\right)\ket{\rho(t)}+\hat{D}\ket*{\rho(t)}
\end{equation}
where $\hat{D}$ is the operator induced by $\mc{D}$. We can therefore define the generator of the time-evolution as $\hat{H}_{\mr{dis}}$, which we call the Hamiltonian in the following (instead of the original model $\hat{H}(t)$) in Eq.\,\eqref{eq:H}),
\begin{equation}
    \hat{H}_{\mr{dis}}\coloneqq-g\sum_{i=1}^L\sum_{\mu\not\in\mf{s}}\left(\hat{S}^+_{i,\mu\mu}+\hat{S}^-_{i,\mu\mu}\right)+\hat{D}.
\end{equation}

We represent each spin operator $\hat{S}^\pm_{i,\mu\nu}$ by the Schwinger bosons via Eq.\,\eqref{eq:Schwinger},
\begin{equation*}
    \hat{S}^\pm_{i,\mu\nu}=\hat{a}_{i\mu\pm}^\dag\hat{a}_{i\nu\pm},\,\,\sum_{\mu=1}^q\hat{a}_{i\mu\pm}^\dag\hat{a}_{i\mu\pm}=\kappa q.
\end{equation*}
This mapping makes the Hamiltonian $\hat{H}_{\mr{dis}}$ a bosonic interacting model. For example, the term corresponding to the first term of $\mc{D}$ in Eq.\,\eqref{eq:dissipator} is written as,
\begin{equation}\label{eq:dissipator Schwinger}
    \begin{split}
        &\frac{2J^2}{q^3}\sum_{i=1}^L\sum_{(\mu,\rho)\not\in\mf{s}^2}\sum_{(\nu,\sigma)\not\in\mf{s}^2}\left[\hat{S}^+_{i,\mu\nu}\hat{S}^-_{i,\mu\nu}\hat{S}^+_{i+1,\rho\sigma}\hat{S}^-_{i+1,\rho\sigma}-\frac{1}{2}\sum_{\alpha=\pm}\left(\hat{S}^\alpha_{i,\mu\nu}\hat{S}^\alpha_{i,\nu\mu}\hat{S}^\alpha_{i+1,\rho\sigma}\hat{S}^\alpha_{i+1,\sigma\rho}\right)\right]\\
        &=\frac{2J^2}{q^3}\sum_{i=1}^L\sum_{(\mu,\rho)\not\in\mf{s}^2}\sum_{(\nu,\sigma)\not\in\mf{s}^2}\left[\big(\hat{a}_{i\mu+}^\dag\hat{a}^\dag_{i\mu-}\big)\big(\hat{a}_{i\nu-}\hat{a}_{i\nu+}\big)\big(\hat{a}^\dag_{i+1\sigma+}\hat{a}_{i+1\sigma-}^\dag\big)\big(\hat{a}_{i+1\mu-}\hat{a}_{i+1\sigma+}\big)\right]\\
        &-\frac{J^2}{q^3}\sum_{i=1}^L\sum_{\alpha=\pm}\sum_{(\mu,\rho)\not\in\mf{s}^2}\sum_{(\nu,\sigma)\not\in\mf{s}^2}\big(\hat{a}_{i\mu\alpha}^\dag\hat{a}_{i\nu\alpha}\hat{a}_{i\nu\alpha}^\dag\hat{a}_{i\mu\alpha}\big)\big(\hat{a}_{i+1\rho\alpha}^\dag\hat{a}_{i+1\sigma\alpha}\hat{a}_{i+1\sigma\alpha}^\dag\hat{a}_{i\rho\alpha}\big).
    \end{split}
\end{equation}

Note that the Schwinger boson representation introduces an emergent $U(1)\times U(1)$ gauge symmetry, 
\begin{equation}
    \hat{a}_{i\mu\alpha}\mapsto e^{i\theta_\alpha}\hat{a}_{i\mu\alpha},\,\,\alpha=\pm.
\end{equation}

Consequently, the physical state should be invariant under the following gauge transformation,
\begin{equation}\label{eq:gauge transformation}
    \begin{split}
        &\hat{U}_i^\alpha\coloneqq\exp\left[i\theta\Big(\sum_{\mu=1}^q\hat{a}_{i\mu\alpha}^\dag\hat{a}_{i\mu\alpha}-\kappa q\Big)\right]\\
        &\hat{U}_i^\alpha\ket{\mr{phys}}=\ket{\mr{phys}}.
    \end{split}
\end{equation} 

\subsection{The $\mf{su}(1,1)$ algebra in the large-$q$ limit}
In the large-$q$ limit, the leading term in Eq.\,\eqref{eq:dissipator Schwinger} consists only of the following combination of the Schwinger boson operators,
\begin{equation}\label{eq:su(1,1)+2}\begin{split}
    \hat{M}_i^z&\coloneqq\frac{1}{2}\sum_{\mu\in\mf{s}}\Big(\hat{a}_{i\mu+}^\dag\hat{a}_{i\mu+}+\hat{a}^\dag_{i\mu-}\hat{a}_{i\mu-}+1\Big),\,\,\hat{M}_i^+\coloneqq\sum_{\mu\in\mf{s}}\hat{a}^\dag_{i\mu+}\hat{a}^\dag_{i\mu-},\,\,
    \hat{M}_i^-=\big(\hat{M}_i^+\big)^\dag\\
    \hat{L}_i^z&\coloneqq\frac{1}{2}\sum_{\mu\not\in\mf{s}}\Big(\hat{a}_{i\mu+}^\dag\hat{a}_{i\mu+}+\hat{a}^\dag_{i\mu-}\hat{a}_{i\mu-}+1\Big),\,\,\hat{L}_i^+\coloneqq\sum_{\mu\not\in\mf{s}}\hat{a}^\dag_{i\mu+}\hat{a}^\dag_{i\mu-},\,\,
    \hat{L}_i^-=\big(\hat{L}_i^+\big)^\dag\\
    \hat{C}_i^M&\coloneqq\sum_{\mu\in\mf{s}}\Big(\hat{a}_{i\mu+}^\dag\hat{a}_{i\mu+}-\hat{a}_{i\mu-}^\dag\hat{a}_{i\mu-}\Big),\,\,\hat{C}_i^L\coloneqq\sum_{\mu\not\in\mf{s}}\Big(\hat{a}_{i\mu+}^\dag\hat{a}_{i\mu+}-\hat{a}_{i\mu-}^\dag\hat{a}_{i\mu-}\Big)
\end{split}
\end{equation}
i.e., Eq.\,\eqref{eq:dissipator Schwinger} is expressed as a polynomial of these operators in this limit. Similarly, one can check that these operators indeed constitute the whole Hamiltonian $\hat{H}_{\mr{dis}}$. Just like the two-photon system in quantum optics, they satisfy the following commutation relation,
\begin{equation}\label{eq:su(1,1)}
    \begin{split}
        \big[\hat{M}^z_i,\hat{M}^\pm_j\big]&=\pm\delta_{ij}\hat{M}_i^\pm,\,\,[\hat{M}^+_i,\hat{M}^-_j]=-2\delta_{ij}\hat{M}_i^z\\
        \big[\hat{L}^z_i,\hat{L}^\pm_j\big]&=\pm\delta_{ij}\hat{L}_i^\pm,\,\,[\hat{L}^+_i,\hat{L}^-_j]=-2\delta_{ij}\hat{L}_i^z,
    \end{split}
\end{equation}
and any other pair of operators commute. This algebra forms the $\mf{su}(1,1)^{\oplus2}\oplus\mb{R}^2$ (at each site), where $\{\hat{M}_i^z,\hat{M}_i^\pm\}$ and $\{\hat{L}_i^z,\hat{L}_i^\pm\}$ are the generators of each $\mf{su}(1,1)$, with the center generated by $\hat{C}_i^M$ and $\hat{C}_i^L$.

We should fix the representation of the algebra Eq.\,\eqref{eq:su(1,1)} for the subsequent analysis. Recall that the initial state is proportional to the identity operator $\bigotimes_i\hat{1}_{SU(q)}$ in the $SU(q)$ spin Hilbert space, where $\hat{1}_{SU(q)}$ is the identity operator. The identity $\hat{1}_{SU(q)}$ is mapped by the Choi–Jamio\l kowski isomorphism to the following state,
\begin{equation}\label{eq:vacuum state}
    \ket*{1_{SU(q)}}=\sum_{\substack{\{n_\mu\}_\mu\\ \sum_{\mu=1}^qn_\mu=\kappa q}}\prod_{\mu=1}^q\frac{\big(\hat{a}_{\mu+}^\dag\hat{a}_{\mu-}^\dag\big)^{n_\mu}}{n_\mu!}\ket*{\mr{vac}}.
\end{equation}  
One can verify this equation by the Schur's lemma. Since $\ket*{1_{SU(q)}}$ is in the kernel of $\hat{a}_{i\mu+}^\dag\hat{a}_{i\mu+}-\hat{a}_{i\mu-}^\dag\hat{a}_{i\mu-}$ for $\forall i$ and $\forall \mu$, we can set the center as $\hat{C}_i^M=\hat{C}_i^L=0$. Note that this is a stronger constraint than the gauge invariance: in the $SU(1,1)$ language, the gauge transformation in Eq.\,\eqref{eq:gauge transformation} is written as 
\begin{equation}
    \hat{U}_i^z\coloneqq\exp\left[i\theta_z\left(\hat{M}_i^z+\hat{L}_i^z-\frac{2\kappa+1}{2}q\right)\right],\,\,\hat{U}_i^c\coloneqq\exp\left[i\theta_c\left(\hat{C}_i^M+\hat{C}_i^L\right)\right].
\end{equation}
Thus, the gauge invariance by $\hat{U}^c_i$ imposes only $\hat{C}^M_i+\hat{C}_i^L=0$. 

Unfortunately, the state $\ket*{1_{SU(q)}}$ does not belong to a single irrep of $SU(1,1)^{\otimes2}$, and a rigorous treatment requires analysis of all the possible irreps. One can, however, prove that physically relevant irreps should be totally symmetric with respect to the permutations of the flavor degrees of freedom $\mu$ within $\mf{s}$ and within its complement $\mf{q}\setminus\mf{s}$. The simplest irrep satisfying this condition is the tensor product of the so-called the discrete series with the Bragmann indices $|\mf{s}|/2$ and $(q-|\mf{s}|)/2$, which we denote by $D_{|\mf{s}|/2}$ and $D_{(q-|\mf{s}|)/2}$, respectively (see Appendix\,\ref{appsec:relevant irreps}). Namely, for the subsequent analysis we have the ansatz,
\begin{equation}
    \ket*{\rho(t)}\in D_{|\mf{s}|/2}\otimes D_{(q-|\mf{s}|)/2}\cap V_{\mr{phys}}.
\end{equation} 
We note that this irrep naturally arises when we consider a simple mean-field approximation (see Appendix\,\ref{appsec:justification}). The subspace $V_{\mr{phys}}$ is the physical subspace satisfying the gauge invariance, 
\begin{equation}\label{eq:gauge invariance}
    \hat{M}_i^z+\hat{L}_i^z=\frac{2\kappa+1}{2}q.
\end{equation}


\subsection{The effective Ginzburg-Landau theory}
Due to the large quantum numbers $|\mf{s}|/2,(q-|\mf{s}|)/2=O(q)$, the dynamics can be studied within the semiclassical approximation. Using the resolution of the identity for $SU(1,1)$, we express the time-evolution of the system by the path integral,
\begin{equation}
    \ket{\rho(t)}=\int D\bm{m}D\bm{l}\delta[G]e^{S[\bm{m},\bm{l}]}\ket{\bm{m}(t),\bm{l}(t)}.
\end{equation}
The state $\ket{\bm{m},\bm{l}}$ is a product of the $SU(1,1)$ spin coherent states,
\begin{equation}\label{eq:coherent state}
    \ket{\bm{m},\bm{l}}=\bigotimes_{i=1}^L\ket{\bm{m}_i}\otimes\ket{\bm{l}_i},
\end{equation}
where $\ket*{\bm{m}_i}$ and $\ket{\bm{l}_i}$ are both the $SU(1,1)$ coherent states (see Appendix\,\ref{appsec:irrep of su(1,1)} for the definition). These states are parametrized by the Bragmann index, which is the analog of the spin quantum number for $SU(2)$, and the unit vector in the upper sheet of the hyperboloid $\mb{H}^2=\{\bm{n};(n^z)^2-(n^x)^2-(n^y)^2=1,n^z>0\}$,
\begin{equation}
    \bm{m}_i=\frac{|\mf{s}|}{2}\begin{pmatrix}
        \sinh\theta_i\cos\phi_i\\
        \sinh\theta_i\sin\phi_i\\
        \cosh\theta_i
    \end{pmatrix}\,(\theta_i>0),\,\,\bm{l}_i=\frac{q-|\mf{s}|}{2}\begin{pmatrix}
        \sinh\Theta_i\cos\Phi_i\\
        \sinh\Theta_i\sin\Phi_i\\
        \cosh\Theta_i
    \end{pmatrix}\,(\Theta_i>0).
\end{equation}
Like the $SU(2)$ spin, the action $S[\bm{m},\bm{l}]$ consists of the topological and the dynamical term,
\begin{equation}\label{eq:action eft}
    \begin{split}
        S[\bm{m},\bm{l}]&=S_{\mr{top}}[\bm{m},\bm{l}]+S_{\mr{dyn}}[\bm{m},\bm{l}]\\
        S_{\mr{top}}[\bm{m},\bm{l}]&=i\frac{|\mf{s}|}{2}\int d\tau\sum_{i=1}^L\left(\cosh\theta_i(\tau)-1\right)\dv{\phi_i(\tau)}{\tau}\\
        &+i\frac{q-|\mf{s}|}{2}\int d\tau\sum_{i=1}^L\left(\cosh\Theta_i(\tau)-1\right)\dv{\Phi_i(\tau)}{\tau}\\
        S_{\mr{dyn}}[\bm{m},\bm{l}]&=\int d\tau\expval{\hat{H}_{\mr{dis}}}{\bm{m}(\tau),\bm{l}(\tau)}.
    \end{split}
\end{equation}
Note that $S_{\mr{dyn}}$ contains the information on the coupling constants $J,V$, and $g$. The delta function $\delta[G]\coloneqq\prod_{i,\tau}\delta[G_i(\tau)]$ imposes the constraint on the particle number Eq.\,\eqref{eq:gauge invariance}, 
\begin{equation}
    G_i(\tau)\coloneqq\frac{|\mf{s}|}{2}\cosh\theta_i(\tau)+\frac{q-|\mf{s}|}{2}\cosh\Theta_i(\tau)-\frac{2\kappa+1}{2}q.
\end{equation}
Thus, one can write $\Theta_i(\tau)$ as a function of $\theta_i(\tau)$,
\begin{equation}\label{eq:particle number each site}
    \Theta_i(\tau)=\mr{arccosh}\left(\frac{2\kappa+1}{1-r}-\frac{r}{1-r}\cosh\theta_i(\tau)\right),
\end{equation}
where $r\coloneqq|\mf{s}|/q$ (the same definition as the RQC model). We also note that the domain of the integral should be the product of the non-negative real axis and the unit circle, i.e., $(\theta_i,\phi_i),(\Theta_i,\Phi_i)\in\mb{R}_{\geq0}\times S^1$. In this region, $\cosh\theta$ is one-to-one and hence $\sinh\theta$ can be uniquely determined once $\cosh\theta$ is given.

One can integrate out $\Theta_i(\tau)$ thanks to the delta function, and write the action $S[\bm{m},\bm{l}]$ as a functional of $\theta_i$ and $\Delta\phi_i\coloneqq\phi_i-\Phi_i$ (up to an irrelevant constant). For example, the topological term becomes
\begin{equation}\label{eq:S_top}
    S_{\mr{top}}[\Delta\phi,\theta]=i\frac{|\mf{s}|}{2}\int d\tau\sum_{i=1}^L\cosh\theta_i(\tau)\dv{\Delta\phi_i(\tau)}{\tau}.
\end{equation}   
This form indicates that the variables $(\Delta\phi,\cosh\theta)$ are canonical coordinates. Note that the integral domain for these fields becomes
\begin{equation}\label{eq:constraint eft}
    (\theta_i(\tau),\Delta\phi(\tau))\in\left[0,\frac{2\kappa+r}{1-r}\right]\times S^1.
\end{equation}

\begin{figure}
    \centering
    \includegraphics[width=.7\textwidth]{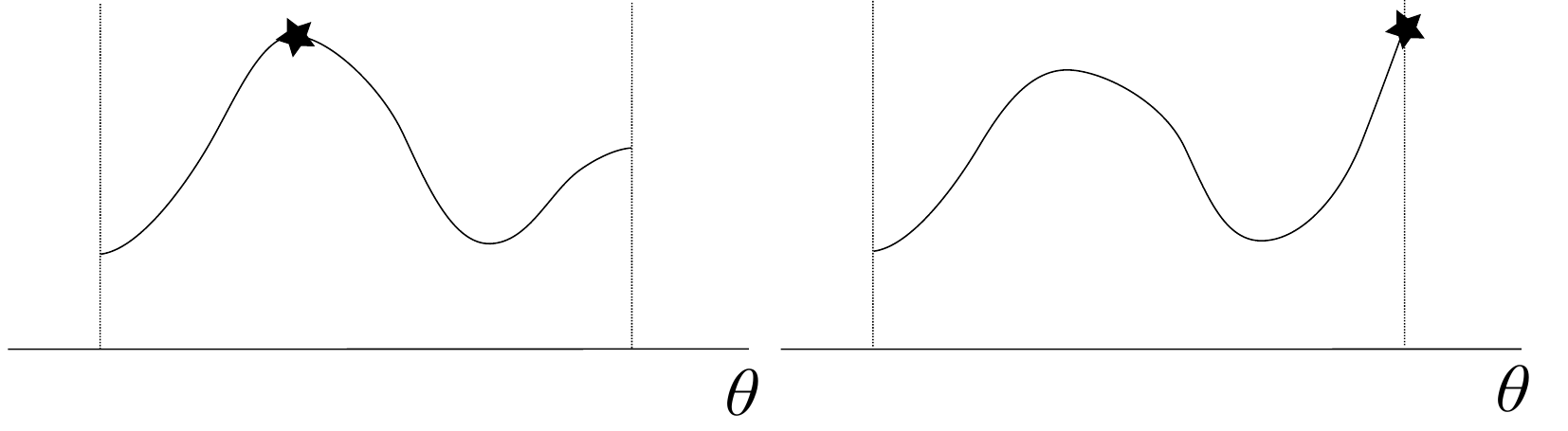}
    \caption{Classification of the optimal configuration in our theory: the left panel describes Case 1, where the optimal point is a local maximum, while Case 2, described in the right panel, the optimal point is not a maximum but at the boundary.}
    \label{fig:phase type}
\end{figure}

The steady state properties are governed by the maximum of the dynamical part $S_{\mr{dyn}}[\Delta\phi,\theta]$ (note the sign of the integrand of the path integral $e^{S}$). To determine the phase structure, we maximize it under the spatio-temporally homogeneous ansatz, 
\begin{equation}\label{eq:mf ansatz}\begin{split}
    \Delta\phi_i(\tau)=\Delta\phi,\,\,\theta_i(\tau)=\theta.
\end{split}
\end{equation}
Before proceeding, we note that the optimal configuration of the energy functional is classified as two types, which we denote as Case 1 and Case 2:
\begin{itemize}
    \item[Case 1] The optimal configuration is a saddle of the action even if we extend the domain of the integral to $\mb{R}_{\geq0}\times S^1$ (see the left panel of Fig.\,\ref{fig:phase type}). Therefore this configuration should solve the Euler-Langrange equation. The structure is therefore qualitatively the same as the conventional Ginzburg-Landau theory.
    \item[Case 2] The optimal configuration is \textit{not} a saddle in the extended domain $\mb{R}_{\geq0}\times S^1$ (see the right panel of Fig.\,\ref{fig:phase type}). In this case, a point at the boundary realizes the optimal configuration, i.e., $\{0, (2\kappa+r)/(1-r)\}\times S^1$ and it does not solve the Euler-Lagrange equation.   
\end{itemize} 

\begin{figure}
    \centering
    \begin{minipage}{0.49\textwidth}
        \includegraphics[width=\linewidth]{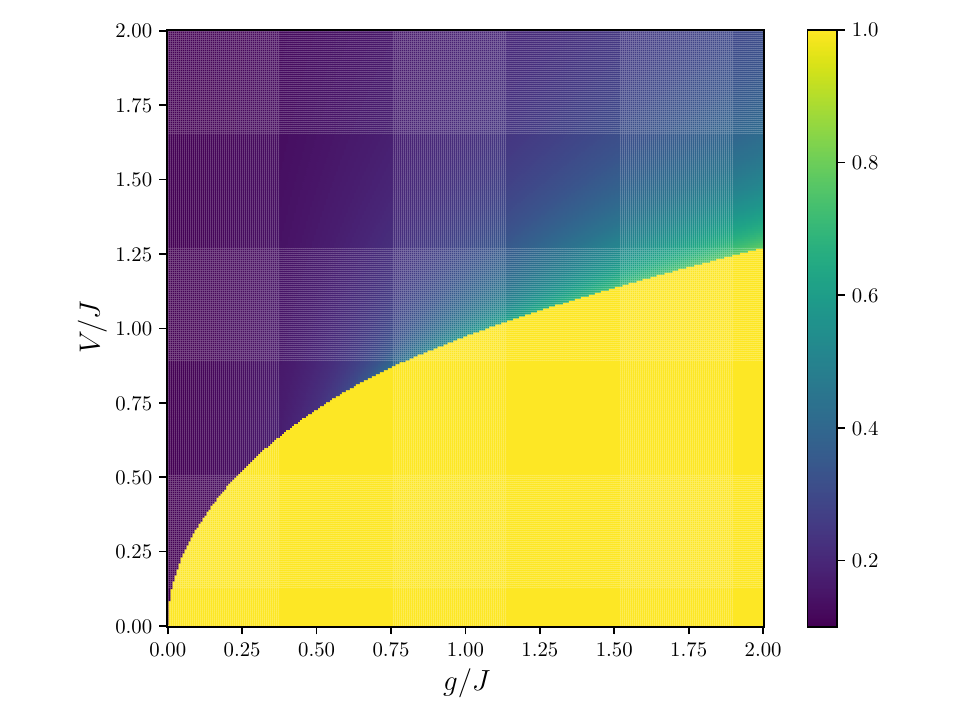}
    \end{minipage}
    \begin{minipage}{0.49\textwidth}
        \includegraphics[width=\linewidth]{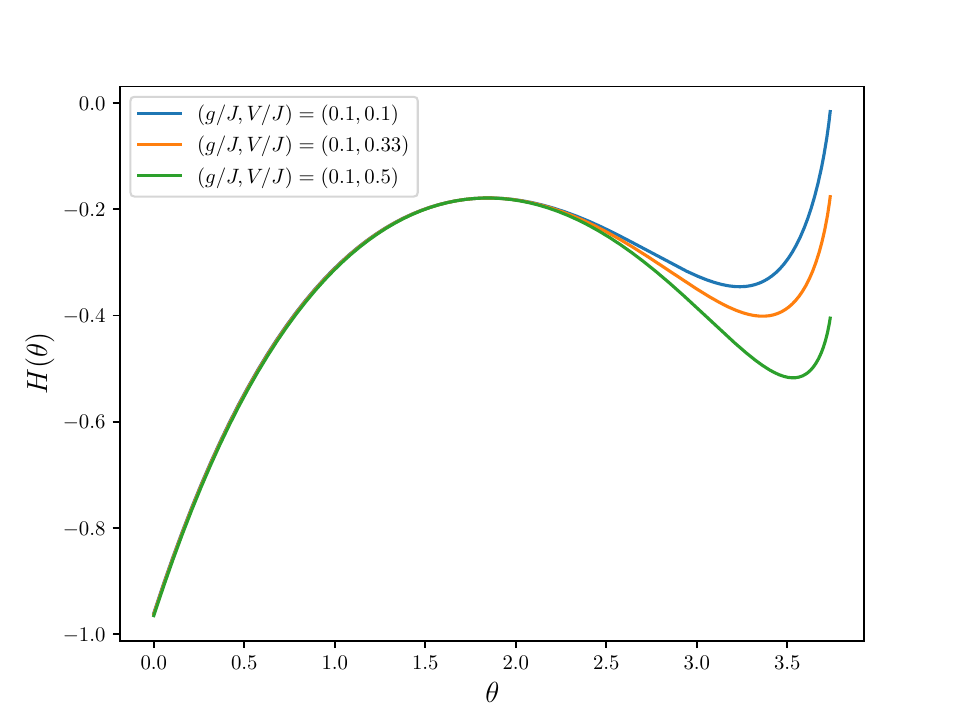}
    \end{minipage}
    \caption{Left panel: the plot of $m^z$ in Eq.\,\eqref{eq:op} for the steady state configuration as a function of $g/J$ and $V/J$. We fix the parameters as $\kappa=1$ and $r=1/10$. The yellow region corresponds to the QMBS phase, and the purple region to the thermal phase.\,\,Right panel: the expectation value of $\hat{H}_{\mr{dis}}$ as a function of the angle $\theta$ ($\Delta\phi$ is set to be zero).}
    \label{fig:diagram EFT}
\end{figure}

We numerically maximize the action Eq.\,\eqref{eq:action eft} under the mean-field ansatz Eq.\,\eqref{eq:mf ansatz}, subject to the constraint Eq.\,\eqref{eq:constraint eft}. In the left panel of Fig.\,\ref{fig:diagram EFT}, we plot the ``$z$-component'' of the $SU(1,1)$ spin in the scar subspace,
\begin{equation}\label{eq:op}
    m^z\coloneqq\frac{1}{2q}\sum_{\mu\in\mf{s}}\expval{\hat{a}_{i\mu+}^\dag\hat{a}_{i\mu+}+\hat{a}_{i\mu-}^\dag\hat{a}_{i\mu-}}=\frac{r}{2}\left(\cosh\theta-1\right),
\end{equation}
which quantifies how much the system's density matrix belongs to the scar subspace, where the average $\expval*{\bullet}$ is taken with respect to the coherent state [Eq.\,\eqref{eq:coherent state}] with the optimal parameter $(\Delta\phi,\theta)$. Figure\,\ref{fig:diagram EFT} exhibits the sharp phase boundary: when $V$ is weak enough, the dissipation of non-scar components induced by the non-Hermitian term $\hat{H}_{\mr{NH}}$ in Eq.\,\eqref{eq:H} favors the scar configuration, resulting in the QMBS phase indicated in the yellow region in Fig.\,\ref{fig:diagram EFT}. On the other hand, if $V$ is strong enough, the entropy gain dominates and the state becomes thermalized. 

The right panel of Fig.\,\ref{fig:diagram EFT} indicates two types of optimal configurations (Case 1 and Case 2) that depend on the coupling constant. When $g/J$ is strong enough, i.e., the system is in the QMBS phase, the optimal $\theta$ is at the boundary of the allowed range $[0,(2\kappa+r)/(1-r)]$, implying the Case 2 configuration. When $V/J$ is instead dominant so the system is thermalized, the optimal configuration satisfies the Euler-Lagrange equation, corresponding to Case 1.

\subsection{Spin wave analysis and finite-$q$ effects}
\begin{figure}
    \centering
    \includegraphics[width=0.7\textwidth]{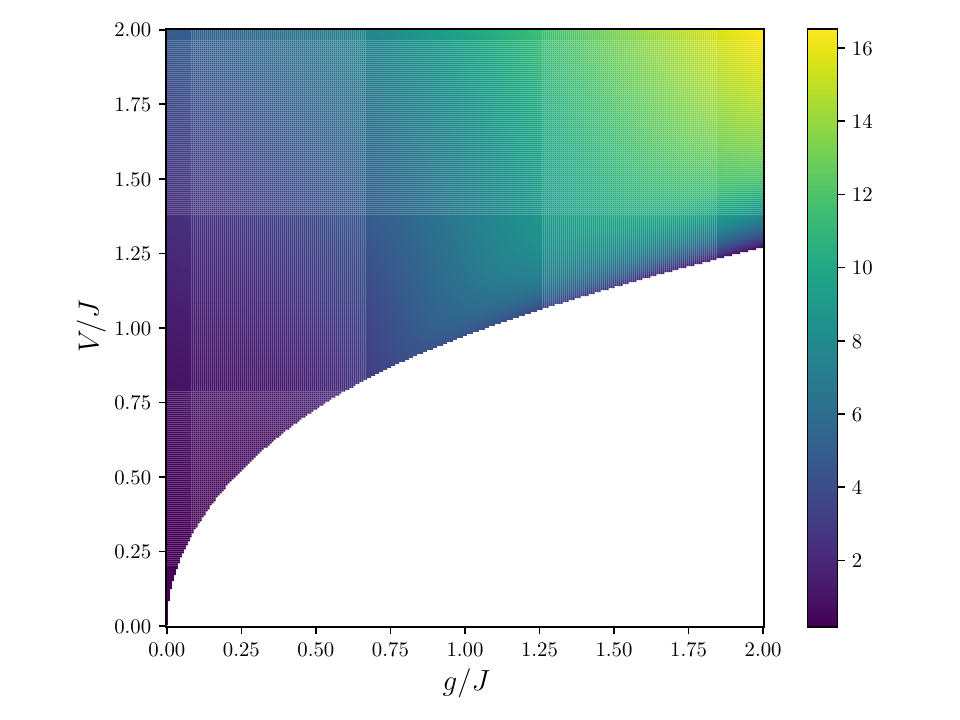}
    \caption{The minimum value of the factor $\min_kt_\vartheta\cos k+m_\vartheta$. The positivity of this factor indicates the stability of the mean-field configuration in Case 1.}
    \label{fig:stability gap}
\end{figure}

The phase transition between the QMBS and the thermal phases is characterized by the discontinuous jump in the ``magnetization'' $m^z$. We confirm the nature of this transition by studying the stability of the mean-field configuration. We define the fluctuating variables $\varphi_i$ and $\vartheta_i$ as 
\begin{equation}
    \Delta\phi_i=\Delta\phi+\varphi_i,\,\,\theta_i=\theta+\vartheta_i,
\end{equation}
where $\Delta\phi$ and $\theta$ are the mean-field value. Expanding the original action with respect to these variables up to the second order, we obtain the effective action $S_{\mr{eff}}$ of the fluctuations,
\begin{equation}
    \begin{split}
    S_{\mr{eff}}&=S_{\mr{quad}}+S_J\\
    S_{\mr{quad}}&\coloneqq i\frac{|\mf{s}|}{2}\int d\tau\sum_{i=1}^L\vartheta_i\dv{\varphi_i}{\tau}-q\int d\tau\sum_{i=1}^L\Big(t_\vartheta\vartheta_i\vartheta_{i+1}+m_\vartheta\vartheta_i^2+t_\varphi\varphi_i\varphi_{i+1}+m_\varphi\varphi_i^2\Big)\\
    S_J&\coloneqq q\int d\tau\sum_{i=1}^LJ_\vartheta\vartheta_i,
    \end{split}
\end{equation}
where the hopping parameters $t_\vartheta$ and $t_\varphi$, the masses $m_\vartheta$ and $m_\varphi$, and the ``source'' $J_\vartheta$ depend on the mean-field value $\theta$. The fields $\varphi$ and $\vartheta$ are decoupled within the Gaussian level and consequently $S_{\mr{quad}}$ is formally identical to the Euclidean theory of a harmonic chain. 
Therefore $S_{\mr{quad}}$ is diagonal in the momentum space, and for each $k$ we can define the simple harmonic oscillator with the mass $m_k$ and the dispersion $\omega_k$, 
\begin{equation}\label{eq:dispersion}
    \begin{split}
        m_k&=\frac{4}{r}\frac{1}{t_\vartheta\cos k+m_\vartheta}\\
    \omega_k&=\frac{4}{r}\sqrt{(t_\vartheta\cos k+m_\vartheta)(t_\varphi\cos k+m_\varphi)}.
    \end{split}
\end{equation}
This indicates that the system is stable if $t_\vartheta\cos k+m_\vartheta$ and $t_\varphi\cos k+m_\varphi$ are both non-negative. One can prove that the latter part $t_\varphi\cos k+m_\varphi$ is always positive real. We also confirm numerically the positivity of the first term $t_\vartheta\cos k+m_\vartheta$ (see Fig.\,\ref{fig:stability gap} and Appendix\,\ref{appsec:RPA}). Therefore, the dispersion is always positive real, implying the stability of the quadratic term and hence the first-order transition between the thermal and the QMBS phase. Note that the optimal $(\Delta\phi,\theta)$ means non-negative $J_\vartheta$ (i.e., $J_\vartheta\geq0$). 

When the saddle point is categorized as Case 1, i.e., the optimal $(\Delta\phi,\theta)$ is a saddle point in the extended domain $\mb{R}_{\geq0}\times S^1$, the source term $S_J$ should vanish because the fields satisfy the Euler-Lagrange equation. Therefore, the stability of $S_{\mr{quad}}$ immediately implies the stability of the system. After integrating the fields $\varphi$ and $\vartheta$ by the formula of the imaginary time propagator of the harmonic oscillator, we obtain the leading correction to the mean-field action,
\begin{equation}
    S\approx S_{\mr{mf}}-\frac{1}{2}\sum_k\left(\ln\left(\frac{4\pi}{|\mf{s}|}\sinh\omega_kT\right)-\ln m_k\omega_k\right),
\end{equation} 
where $S_{\mr{mf}}$ is the mean-field action, and $T$ is the time after quench. This is a $1/q$ correction, as the dominant term in the correction is $\sum_k\ln\sinh\omega_kT=O(LT)$, while $S_{\mr{saddle}}=O(qLT)$. 

When the saddle point belongs to Case 2, i.e., the optimal $(\Delta\phi,\theta)$ is at the boundary of the domain, which is realized in the QMBS phase, the source term does not vanish and rather becomes the leading correction. In this case, we can ignore the quadratic part $S_{\mr{quad}}$ as a higher order term. Since $\vartheta_i$ explores only the negative real line, i.e., $\vartheta_i\leq0$ (note $\cosh(\theta+\vartheta_i)\leq\cosh\theta$), we can integrate this field as the half-Lorenzian. The path integral is thus approximated as,
\begin{equation}
    \int D\Delta\phi D\theta e^{S[\Delta\phi,\theta]}\approx e^{S_{\mr{mf}}}\prod_{i=1}^L\int^{\vartheta_i(t)=0}_{\vartheta_i(t)=\vartheta_{\mr{min}}}D\vartheta_ie^{S_J}=e^{S_{\mr{mf}}}\prod_{i=1}^L\left(\frac{1-e^{\vartheta_{\mr{min}}}}{qJ_\vartheta}\right)^T,
\end{equation}
where $\vartheta_{\mr{min}}\coloneqq-(2\kappa+r)/(1-r)$. This expression indicates that the leading correction to the action is,
\begin{equation}
    S\approx S_{\mr{mf}}-LT\ln(\frac{qJ_\vartheta}{1-e^{\vartheta_{\mr{min}}}}),
\end{equation}   
which is a $\ln q/q$ correction. 

The phase dependent leading order correction to the mean-field action implies the modification to the mean-field phase diagram: since the correction is always parametrically larger for the QMBS phase, the critical coupling $g_c(q,V,J)$ as a function of $q$, $J$ and $V$ should be modified as,
\begin{equation}
    g_c(q,J,V)=g_\infty(J,V)+c(J,V)\frac{\ln q}{q}+O(1/q),
\end{equation} 
where $g_\infty(J,V)$ is the critical coupling for the large-$q$ calculation, and $c(J,V)>0$ is a positive function depending only on $J$ and $V$. 

\section{Numerical result for the spin-1 XY model}
\label{sec:numerics on XY}
In this section, we present exact diagonalization results of the spin-1 XY model discussed in Sec.\,\ref{sec:motivation} for a small system size. This allows us to directly access to the steady state, in contrast to TEBD, which cannot provide reliable approximations for long-time dynamics. The Hamiltonian is given by,
\begin{equation*}\begin{split}
    \hat{H}&=\hat{H}_0+\hat{H}_{\mr{NH}}+\hat{V}\\
    \hat{H}_0&=J\sum_{i=1}^L\left(\hat{S}_i^x\hat{S}_{i+1}^x+\hat{S}_i^y\hat{S}_{i+1}^y\right)+\sum_{i=1}^L\left(h\hat{S}_i^z+D(\hat{S}_i^z)^2\right)\\
    \hat{H}_{\mr{NH}}&=-ig\sum_{i=1}^L\left(1-(\hat{S}_i^z)^2\right)\\
    \hat{V}&=V\sum_{i=1}^L\hat{S}_i^x.    
\end{split}
\end{equation*}
As discussed in Sec.\,\ref{sec:motivation}, we employ the operator $\hat{O}\coloneqq L^{-1}\sum_{i=1}^L(\hat{S}_i^z)^2$ as a measure for ergodicity breaking. This operator can distinguish the scar states from other thermal states, as $\ket{S_n}$ for the clean XY model satisfy $\expval*{\hat{O}}{S_n}=1$, while the Gibbs ensemble must satisfy $\mr{tr}\hat{\rho}_{\mr{Gibbs}}\hat{O}<1$. 

\begin{figure}
\begin{minipage}{0.49\textwidth}
    \centering
    \includegraphics[width=0.9\linewidth]{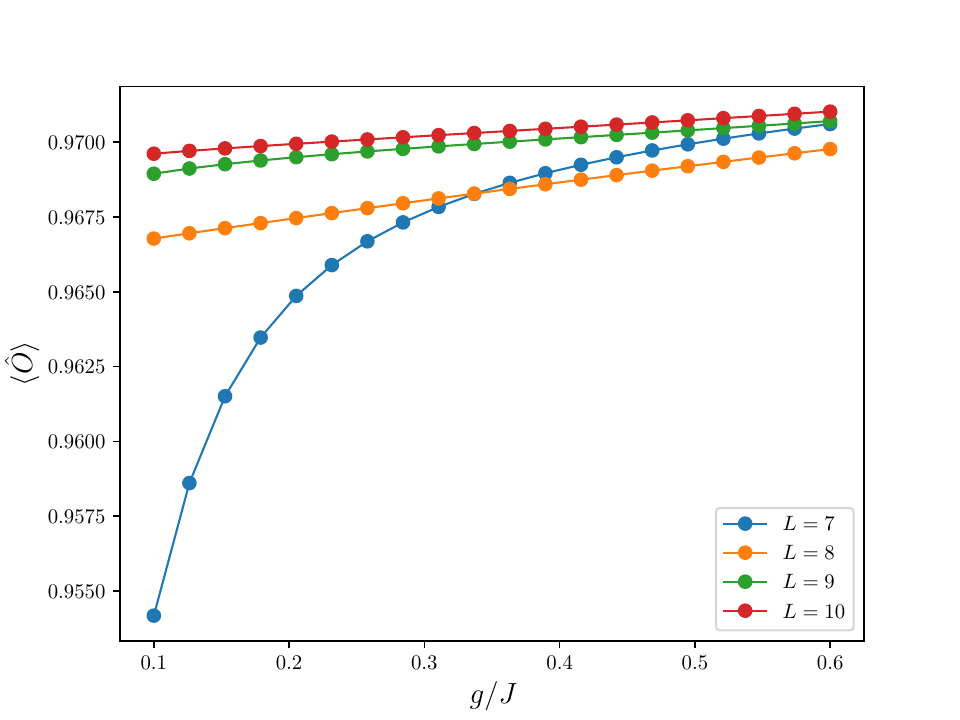}
\end{minipage}
\begin{minipage}{0.49\textwidth}
    \centering
    \includegraphics[width=0.9\linewidth]{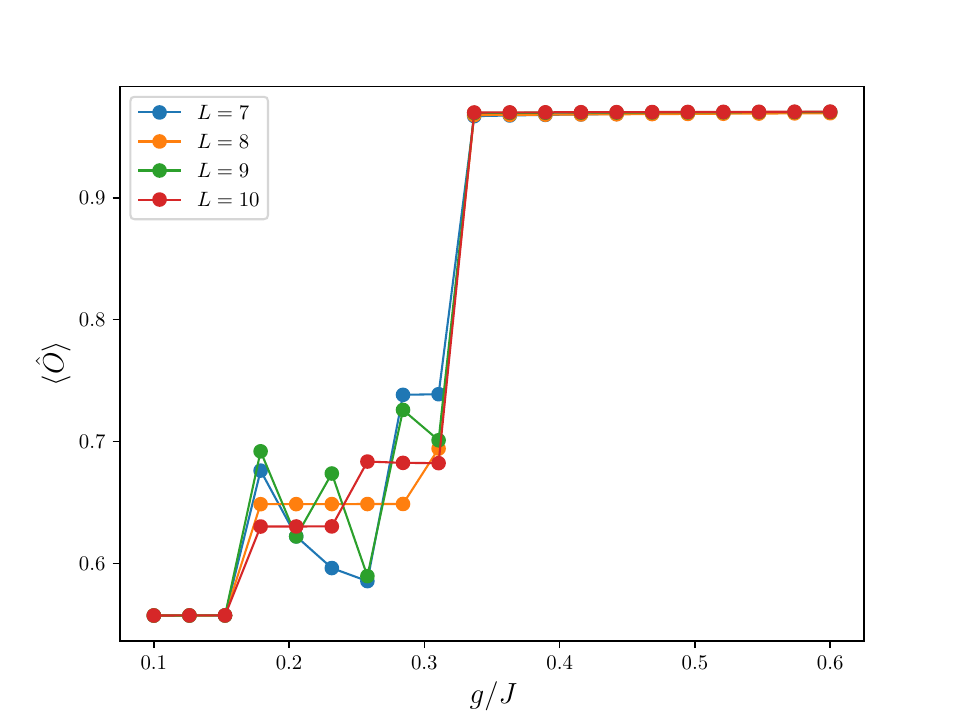}
\end{minipage}
    \caption{The expectation value of the order parameter $\hat{O}$ as a function of the strength of the non-Hermitian term $g$. Left panel: the perturbation is Hermitian, i.e., $V$ is real (System 1). We set them as $J=h=D=V=1$. Right panel: The perturbation is non-Hermitian, and $V$ is complex (System 2). We set them as $J=h=D=1,$ and $V=1+0.1i$.}
    \label{fig:orderparam1}
\end{figure}
We calculate the expectation value of the order parameter $\hat{O}$ with respect to the steady state (the corresponding eigenstate of the energy with the largest imaginary part). To analyze the competition between the non-Hermitian term and the perturbation carefully, we consider the following two systems,
\begin{itemize}
    \item[System 1] The coupling constant $V$ in the perturbation is real. This setting corresponds to the closed system with the engineered dissipation by $\hat{H}_{\mr{NH}}$.
    \item[System 2] The coupling constant $V$ in the perturbation is complex. In this case, the perturbation represents dissipation, as we considered for the RQC model and its semi-classical theory. While strictly speaking we have to consider the Lindblad-type operator for a more rigorous treatment on dissipation, we find that this simplified setting already exhibits the essential features of the phase transition between the thermal and the QMBS phase.
\end{itemize}

\begin{figure}
    \begin{minipage}{0.49\textwidth}
        \centering
        \includegraphics[width=0.9\linewidth]{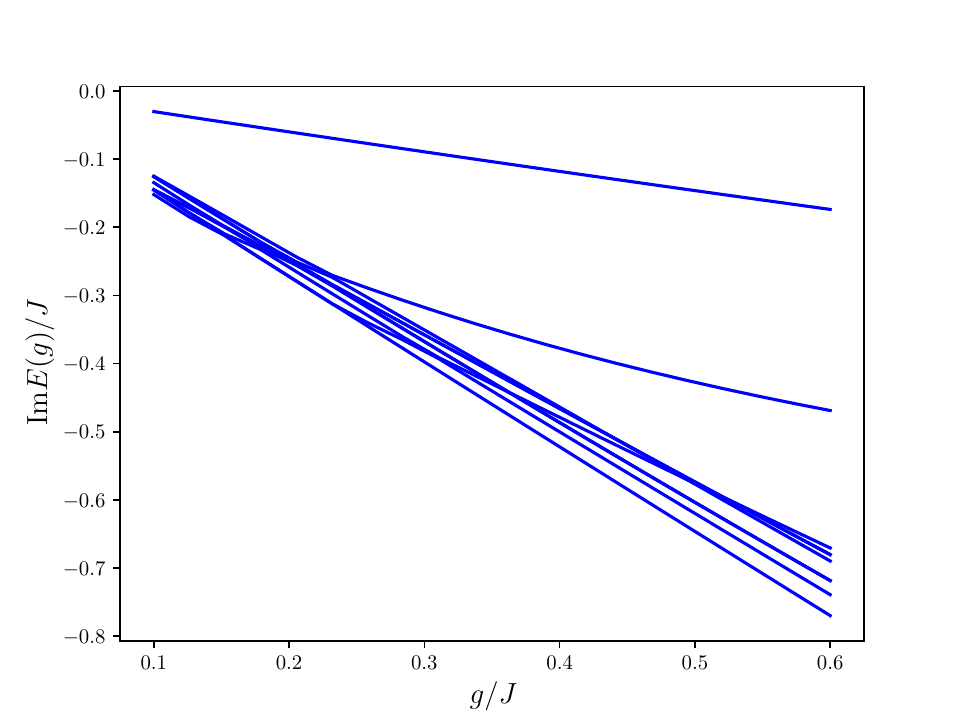}
    \end{minipage}
    \begin{minipage}{0.49\textwidth}
        \centering
        \includegraphics[width=0.9\linewidth]{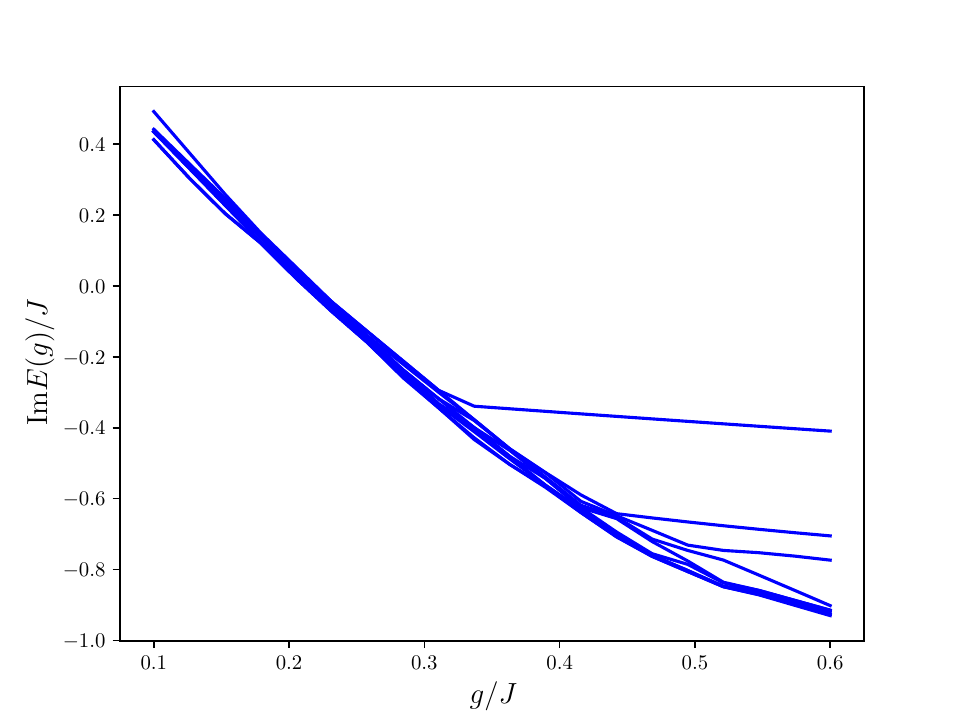}
    \end{minipage}
    \centering
    \includegraphics[width=0.95\linewidth]{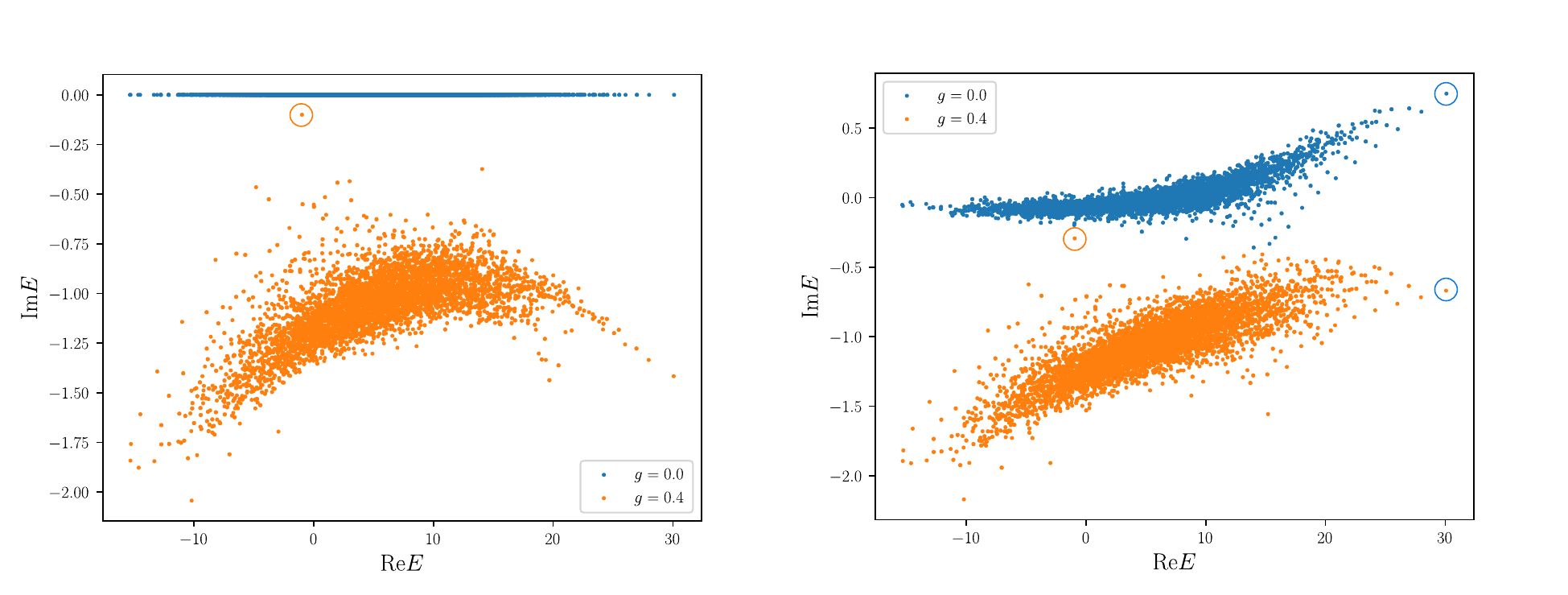}
    \caption{Top panels: The imaginary part of 10 energy eigenvalues as a function of the strength of the non-Hermitian term $g$. Top left: When the other terms are Hermite (System 1), there always exists a gap separating the steady state and other eigenstates. Top Right: When the non-Hermitian perturbation is considered (System 2), in the thermal phase the imaginary part of the energy is highly degenerate. This degeneracy becomes lifted as $g$ is increased. Bottom panels: The many-body spectrum for $L=8$. Bottom left: When the other terms are Hermite (System 1), a small non-Hermitian term $\hat{H}_{\mr{NH}}$ is sufficient to stabilize the scar state. The orange marked circle indicates the scarred steady state. Bottom right: When the non-Hermitian perturbation is added (System 2), there is a genuine competition between the scar state (the orange circle) and the thermal state (the blue circle).}
    \label{fig:level}
\end{figure}

For System 1, even a weak non-Hermitian term drives the order parameter close to unity, signaling entry into the QMBS phase (the left panel of Fig.\,\ref{fig:orderparam1}). We examine values of $g$ as small as $g/J=10^{-4}$, and find the system is already in the QMBS phase. Based on this observation, we conjecture that an infinitesimal non-Hermitian term is sufficient to induce the QMBS phase for System 1. By contrast, System 2 exhibits a clear first-order jump in the order parameter at the transition point (the right panel of Fig.\,\ref{fig:orderparam1}). This behavior is also reflected in the many-body spectrum in Fig.\,\ref{fig:level}: the non-Hermitian term $H_{\mr{NH}}$ generally shift the imaginary part $\Im E$ of all energy levels downward. System 1, when $g=0$ (so $H_{\mr{NH}}=0$), the entire spectrum lies on the real axis. As $g$ increases, every level's imaginary part decreases, yet the scarred steady state are largely unaffected, since it is nearly annihilated by $\hat{H}_{\mr{NH}}$. System 2 follows a similar pattern, but with one key difference: $\hat{H}_0+\hat{V}$ is already non-Hermitian at $g=0$, so the initial energy levels span the full complex plane. Therefore, even when $g=0$, there already exists the steady state in the thermal phase with the largest $\Im E$. As $g$ grows, the imaginary part of the energy of this steady state shrinks more rapidly than that of the scarred steady state. Beyond the critical $g$, however, the scarred state overtakes it in $\Im E$, triggering the phase transition.  

\begin{figure}
    \begin{minipage}{0.49\textwidth}
        \centering
        \includegraphics[width=\linewidth]{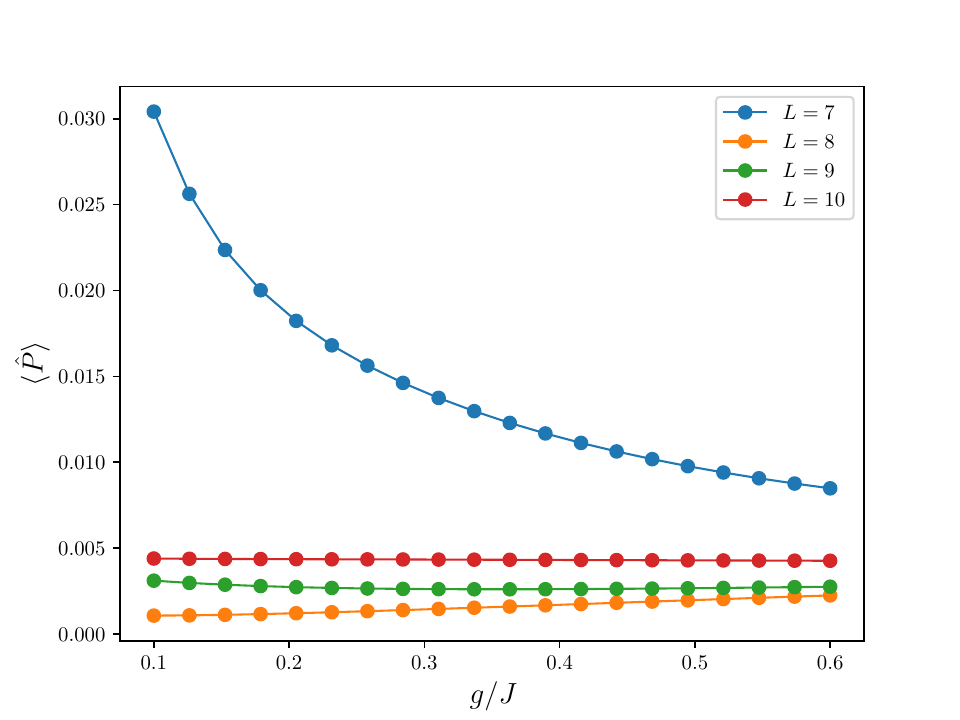}
    \end{minipage}
    \begin{minipage}{0.49\textwidth}
        \centering
        \includegraphics[width=\linewidth]{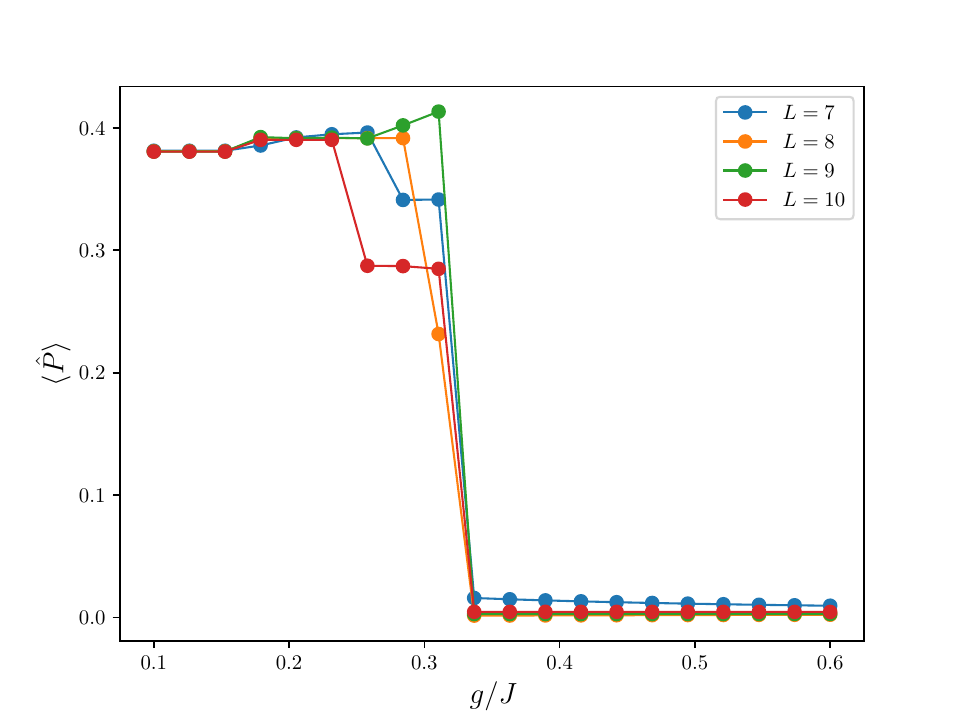}
    \end{minipage}
    \caption{The expectation value of $\hat{P}^{\mr{XY}}_{i,i+1}$ in Eq.\,\eqref{eq:xy annihilator}.}
    \label{fig:orderparam2}
\end{figure}

Finally, we measure how much the steady state look close to one of the scar states $\ket{S_n}$ in the clean spin-1 XY model, by computing the expectation value of $\hat{P}^{\mr{XY}}_{i,i+1}$ in Eq.\,\eqref{eq:xy annihilator} which annihilates $\ket*{S_n}$. Figure\,\eqref{fig:orderparam2} indicates that if all the other terms are Hermitian, a small non-Hermitian term can force the steady state to be almost annihilated by $\hat{P}^{\mr{XY}}_{i,i+1}$. If the non-Hermitian perturbation is allowed, we clearly observe the transition and in the QMBS phase the expectation value of $\hat{P}^{\mr{XY}}_{i,i+1}$ becomes nearly zero. This implies that in the QMBS phase, the steady state should be locally very close to the scar states. 

\section{Conclusion}
\label{sec:conclusion}
This paper demonstrates the existence of a novel phase of matter---the QMBS phase---in non-Hermitian dynamics. The transition can be understood as the non-equilibrium analog of the first-order phase transition, induced by the competition between the entropy gain of the thermal state and the selective stabilization of the scar states. 
We verify the existence of the QMBS phase by three systems, the RQC, the $SU(q)$ spin chain with large spin quantum number, and the spin-1 XY model. 

The RQC model is mapped to the two dimensional classical statistical mechanics model, and the corresponding partition function is evaluated by rotating the quantization axis, which renders the chain of quantum gates a simple transfer matrix. The transfer matrix is approximated by both the mean-field approximation and DMRG, which agree even quantitatively, showing the existence of the phase transition between the QMBS and the thermal phase. These approximations are supplemented by the real space RG, which also confirms the existence of these phases and the first order phase transition between them.

The $SU(q)$ spin chain, a continuous time version of the RQC model, is analyzed in the semiclassical limit. In this limit, the dynamics are effectively generated by the emergent classical $SU(1,1)$ spin, which allows us to construct the non-equilibrium analog of free energy functional. The mean-field theory exhibits the QMBS phase and the thermal phase separated by the first-order transition. We also find that the finite-$q$ correction is qualitatively different for these phases because of their peculiar energy landscapes, potentially shifting the phase boundary unidirectionally.  

Finally, we numerically study the spin-1 XY model with the perturbation. When the perturbation is Hermite, even a small non-Hermitian term can drive the system into the QMBS phase, while there exists a critical coupling constant separating two phases if the perturbation is also non-Hermitian. Intuitively, this behavior is understood as the shift in the complex energy by the non-Hermitian drive that forces the whole energy spectrum to have smaller imaginary part in their energy. The steady state in the QMBS phase is less affected by this effect since it is almost annihilated by the non-Hermitian driving.  

A number of follow-up questions should be discussed: in the simplest QMBS phase discussed in this paper, the transition is discontinuous. It is an interesting question whether this transition must be first-order or continuous transitions are possible. For such continuous transitions, one could discuss their universality classes. Furthermore, given that the essence of the QMBS phase is to stabilize otherwise fragile states by selective measurements, could it be possible to protect other exotic states, such as topologically non-trivial wavefunctions? On the practical side, following a concrete experimental setup for the stabilization of QMBS by ``Fock skin effect''~\cite{PhysRevLett.133.216601}, quantum circuit realizations to observe the QMBS phase may be possible, at least by extensive post-selections or by fault-tolerant quantum computers (FTQC). However, like MIPT, constructing experimental setups that require only noisy intermediate-scale quantum (NISQ) devices without extensive post-selection is an important open question.

\section*{Acknowledgments}
We thank Y.\,Hidaka for stimulating discussions. KO is supported by Grant No.\,204801021001 of the SNSF.

\appendix
\section{Brief review of the scar states of the PXP model}
\label{appendix:pxp}
In this section, we provide a succinct summary of the results in Refs.\,\cite{PhysRevA.107.023318,PhysRevB.108.054412}. The deformed PXP model $\hat{H}'$ still commutes with the Rydberg projector $\hat{P}_{\mr{Ryd}}$,
\begin{equation}
    \hat{H}'_{\mr{PXP}}\hat{P}_{\mr{Ryd}}=\hat{P}_{\mr{Ryd}}\hat{H}'_{\mr{PXP}}.
\end{equation}
On the right-hand side, we can add or subtract an arbitrary term $\hat{V}$ as long as it is annihilated by $\hat{P}_{\mr{Ryd}}$ from the left, i.e., $\hat{P}_{\mr{Ryd}}\hat{V}=0$. Our aim is to engineer a proper $\hat{V}$ so that $\hat{H}'_{\mr{PXP}}+\hat{V}$ becomes the SM form [Eq.\,\eqref{eq:SM general} in the main text]. Such a $\hat{V}$ turns out to be the following,
\begin{equation}\begin{split}
    \hat{V}&=\sum_{i=1}^L\left(\frac{3}{4}\ket{\uparrow\uparrow}\left(\bra*{\uparrow\downarrow}+\bra*{\downarrow\uparrow}\right)_{i,i+1}-\dyad{\uparrow\uparrow\uparrow}{\downarrow\downarrow\downarrow}_{i-1,i,i+1}\right)\\
    &+\frac{1}{4}\sum_{i=1}^L\left(\dyad*{\uparrow\uparrow\downarrow\downarrow}{\uparrow\downarrow\downarrow\downarrow}+\dyad*{\downarrow\downarrow\uparrow\uparrow}{\downarrow\downarrow\downarrow\uparrow}\right)_{i-1,i,i+1,i+2}.
    \end{split}
\end{equation}
Indeed, we find,
\begin{equation}\begin{split}
    \hat{H}'_{\mr{PXP}}+\hat{V}&=\sum_{i=1}^L\left(\hat{W}_i-\frac{1}{2}\left(\ket*{\uparrow\downarrow}+\ket*{\downarrow\uparrow}\right)\left(\bra{\uparrow\uparrow}-\frac{1}{2}\bra*{\downarrow\downarrow}\right)_{i,i+1}\right)\\
    &-\frac{1}{2}\sum_{i=1}^L\left(\ket*{\uparrow\downarrow\downarrow}\left(\bra*{\uparrow\uparrow\downarrow}-\frac{1}{2}\bra*{\downarrow\downarrow\downarrow}\right)+\ket*{\downarrow\downarrow\uparrow}\left(\bra*{\downarrow\uparrow\uparrow}-\frac{1}{2}\bra*{\downarrow\downarrow\downarrow}\right)\right)_{i-1,i,i+1},
    \end{split}
\end{equation}
where $\hat{W}\coloneq\hat{\sigma}^+/2+\hat{\sigma}^-$. The eigenstate of the single-body term $\hat{W}$ is 
\begin{equation}
    \begin{split}
        \hat{W}\ket*{\widetilde{+}}&=\frac{1}{\sqrt{2}}\ket*{\widetilde{+}},\,\,\ket*{\widetilde{+}}\coloneq\sqrt{\frac{1}{3}}\ket{\uparrow}+\sqrt{\frac{2}{3}}\ket{\downarrow}\\
        \hat{W}\ket*{\widetilde{-}}&=-\frac{1}{\sqrt{2}}\ket*{\widetilde{-}},\,\,\ket*{\widetilde{-}}\coloneq\sqrt{\frac{1}{3}}\ket{\uparrow}-\sqrt{\frac{2}{3}}\ket{\downarrow}.
    \end{split}
\end{equation}
Any term in the interaction part of $\hat{H}'_{\mr{PXP}}+\hat{V}$ contains the bra vector $\bra*{\uparrow\uparrow}-1/2\bra*{\downarrow\downarrow}$. We can find that the following three ket vectors are all orthogonal to this bra vector,
\begin{equation}
    \ket*{\widetilde{+}\widetilde{+}},\,\,\frac{1}{\sqrt{2}}\left(\ket*{\widetilde{+}\widetilde{-}}-\ket*{\widetilde{-}\widetilde{+}}\right),\,\,\ket*{\widetilde{-}\widetilde{-}}.
\end{equation}
Note that these vectors are also eigenstates of the Zeeman term consisting of $\hat{W}$. Therefore, we find the following exact eigenstates $\ket*{\widetilde{S}_n}$ of $\hat{H}'_{\mr{PXP}}+\hat{V}$,
\begin{equation}
    \ket*{\widetilde{S}_n}=\frac{1}{N_n}\left(\sum_{i=1}^L(-1)^i\dyad*{\widetilde{+}}{\widetilde{-}'}\right)^n\ket*{\widetilde{-}\cdots\widetilde{-}},
\end{equation}
where $\bra*{\widetilde{-}'}$ is the dual of $\ket*{\widetilde{-}}$, i.e., $\innerproduct*{\widetilde{-}'}{\widetilde{-}}=1$ and $\innerproduct*{\widetilde{-}'}{\widetilde{+}}=0$. The connection to the SM structure is now evident.

\section{Derivation of the master equation}
\label{appsec:master equation}
The effective time-evolution is derived in the same manner as the derivation of the quantum master equation. The von Neumann equation is 
\begin{equation}\label{eq:vN eq}
    \dv{\hat{\rho}(t)}{t}=-i\llbracket\hat{H}(t),\hat{\rho}(t)\rrbracket,
\end{equation}
where $\llbracket\hat{A},\hat{B}\rrbracket$ is the generalized commutator for non-Hermitian dynamics,
\begin{equation}
    \llbracket\hat{A},\hat{B}\rrbracket\coloneqq\hat{A}\hat{B}-\hat{B}\hat{A}^\dag.
\end{equation}
We introduce the ``interaction picture'', 
\begin{equation}\label{eq:vN eq int}
    \hat{\rho}^I(t)\coloneqq e^{i\hat{H}_{\mr{NH}}t}\hat{\rho}(t)e^{-i\hat{H}^\dag_{\mr{NH}}t},\,\,\hat{O}^I(t)\coloneq e^{i\hat{H}_{\mr{NH}}t}\hat{O}(t)e^{-i\hat{H}_{\mr{NH}}t}.
\end{equation}
The von Neumann equation now becomes,
\begin{equation}
    \dv{\hat{\rho}^I(t)}{t}=-i\big\llbracket\hat{V}^I(t),\hat{\rho}^I(t)\big\rrbracket,
\end{equation}
where $\hat{V}(t)$ is the interaction term ($\hat{V}(t)=\hat{H}(t)-\hat{H}_{\mr{NH}}$) in the Hamiltonian. 
The formal solution of Eq.\,\eqref{eq:vN eq int} is 
\begin{equation}
    \hat{\rho}^I(t)=\hat{\rho}(0)-\int^td\tau\big\llbracket\hat{V}^I(\tau),\hat{\rho}(0)\big\rrbracket-\int^t\int^\tau d\tau d\tau'\big\llbracket\hat{V}^I(\tau),\big\llbracket\hat{H}(\tau'),\hat{\rho}(\tau')\big\rrbracket\big\rrbracket.
\end{equation}
We take the disorder average (with respect to $J_{ij}(t)$, and $V_{ij}(t)$) and obtain,
\begin{equation}\begin{split}
    \overline{\hat{\rho}^I(t)}&=\hat{\rho}(0)-\int^t\int^\tau d\tau d\tau'\overline{\big\llbracket\hat{V}^I(\tau),\big\llbracket\hat{V}^I(\tau'),\hat{\rho}^I(\tau')\big\rrbracket\big\rrbracket}\\
    &=\hat{\rho}(0)+\int^t d\tau e^{i\hat{H}_{\mr{NH}}\tau}\mc{D}[\overline{\hat{\rho}(\tau)}]e^{-i\hat{H}^\dag_{\mr{NH}}\tau},
\end{split}
\end{equation}
where the super-operator $\mc{D}$ is defined as
\begin{equation}
    \begin{split}
        \mc{D}[\hat{\rho}]&=\frac{2J^2}{q^3}\sum_{i=1}^L\sum_{(\mu,\rho)\not\in\mf{s}^2}\sum_{(\nu,\sigma)\not\in\mf{s}^2}\left[\hat{S}_{i,\mu\nu}\hat{S}_{i+1,\rho\sigma}\hat{\rho}\hat{S}_{i,\nu\mu}\hat{S}_{i+1,\sigma\rho}-\frac{1}{2}\left\{\hat{S}_{i,\mu\nu}\hat{S}_{i,\nu\mu}\hat{S}_{i+1,\rho\sigma}\hat{S}_{i+1,\sigma\rho},\hat{\rho}\right\}\right]\\
        &+\frac{2V^2}{q^3}\sum_{i=1}^L\sum_{\mu\rho\not\in\mf{s}}\sum_{\nu\sigma\in\mf{s}}\left[\hat{S}_{i,\mu\nu}\hat{S}_{i+1,\rho\sigma}\hat{\rho}\hat{S}_{i,\nu\mu}\hat{S}_{i+1,\sigma\rho}-\frac{1}{2}\left\{\hat{S}_{i,\mu\nu}\hat{S}_{i,\nu\mu}\hat{S}_{i+1,\rho\sigma}\hat{S}_{i+1,\sigma\rho},\hat{\rho}\right\}\right]\\
        &+\frac{2V^2}{q^3}\sum_{i=1}^L\sum_{\mu\rho\not\in\mf{s}}\sum_{\nu\sigma\in\mf{s}}\left[\hat{S}_{i,\nu\mu}\hat{S}_{i+1,\sigma\rho}\hat{\rho}\hat{S}_{i,\mu\nu}\hat{S}_{i+1,\rho\sigma}-\frac{1}{2}\left\{\hat{S}_{i,\nu\mu}\hat{S}_{i,\mu\nu}\hat{S}_{i+1,\sigma\rho}\hat{S}_{i+1,\rho\sigma},\hat{\rho}\right\}\right]\\
        &\equiv\sum_{i=1}^L\left(\mc{D}^1_i[\hat{\rho}]+\mc{D}_i^2[\hat{\rho}]+\mc{D}^3_i[\hat{\rho}]\right).
    \end{split}
\end{equation}
Therefore, the averaged density matrix satisfies the quantum master equation,
\begin{equation}
    \dv{\overline{\hat{\rho}(t)}}{t}=-i\Big\llbracket\hat{H}_{\mr{NH}},\overline{\hat{\rho}(t)}\Big\rrbracket+\mc{D}[\overline{\hat{\rho}(t)}].
\end{equation}

By the Choi–Jamio\l kowski isomorphism, the superoperators $\mc{D}_i^\alpha$ are mapped to,
\begin{equation}\label{eq:L^a}
    \begin{split}
        \hat{D}_i^1&=\frac{2J^2}{q^3}\sum_{(\mu,\rho)\not\in\mf{s}^2}\sum_{(\nu,\sigma)\not\in\mf{s}^2}\left[\hat{S}_{i,\mu\nu}^+\hat{S}^-_{i,\mu\nu}\hat{S}^+_{i+1,\rho\sigma}\hat{S}^-_{i+1,\rho\sigma}-\frac{1}{2}\sum_{\alpha=\pm}\hat{S}_{i,\mu\nu}^\alpha\hat{S}_{i,\nu\mu}^\alpha\hat{S}_{i+1,\rho\sigma}^\alpha\hat{S}^\alpha_{i+1,\sigma\rho}\right]\\
        \hat{D}_i^2&=\frac{2V^2}{q^3}\sum_{\mu\rho\not\in\mf{s}}\sum_{\nu\sigma\in\mf{s}}\left[\hat{S}^+_{i,\mu\nu}\hat{S}^-_{i,\mu\nu}\hat{S}^+_{i+1,\rho\sigma}\hat{S}^-_{i+1,\rho\sigma}-\frac{1}{2}\sum_{\alpha=\pm}\hat{S}_{i,\mu\nu}^\alpha\hat{S}^\alpha_{i,\nu\mu}\hat{S}_{i+1,\rho\sigma}^\alpha\hat{S}_{i+1,\sigma\rho}^\alpha\right]\\
        \hat{D}_i^3&=\frac{2V^2}{q^3}\sum_{\mu\rho\not\in\mf{s}}\sum_{\nu\sigma\in\mf{s}}\left[\hat{S}^+_{i,\nu\mu}\hat{S}^-_{i,\nu\mu}\hat{S}_{i+1,\sigma\rho}^+\hat{S}_{i+1,\sigma\rho}^--\frac{1}{2}\sum_{\alpha=\pm}\hat{S}^\alpha_{i,\nu\mu}\hat{S}^\alpha_{i,\mu\nu}\hat{S}^\alpha_{i+1,\sigma\rho}\hat{S}^\alpha_{i+1,\rho\sigma}\right].
    \end{split}
\end{equation} 

In the large-$q$ limit, the leading order of Eq.\,\eqref{eq:L^a} can be expressed as the $SU(1,1)$ spin operators with the center $\hat{C}_i^M=\hat{C}_i^L=0$, 
\begin{equation}\label{eq:D su(1,1)}
    \begin{split}
        &\hat{D}_i^1=\frac{2J^2}{q^3}\left(\hat{J}_i^+\hat{J}_{i+1}^+-\hat{M}_i^+\hat{M}_{i+1}^+\right)\left(\hat{J}_i^-\hat{J}_{i+1}^--\hat{M}_i^-\hat{M}_{i+1}^-\right)\\
        &-\frac{2J^2}{q^3}\left(\kappa^2q^2-\left(\hat{M}_i^z-\frac{|\mf{s}|}{2}\right)\left(\hat{M}_{i+1}^z-\frac{|\mf{s}|}{2}\right)\right)\left((\kappa+1)^2q^2-\left(\hat{M}_i^z+\frac{|\mf{s}|}{2}\right)\left(\hat{M}_{i+1}^z+\frac{|\mf{s}|}{2}\right)\right)\\
        &\hat{D}_i^2=\frac{2V^2}{q^3}\hat{L}_i^+\hat{L}_{i+1}^+\hat{M}_i^-\hat{M}_{i+1}^-\\
        &-\frac{2V^2}{q^3}\left(\hat{L}_i^z-\frac{q-|\mf{s}|}{2}\right)\left(\hat{L}_{i+1}^z-\frac{q-|\mf{s}|}{2}\right)\left(\hat{M}_i^z+\frac{|\mf{s}|}{2}\right)\left(\hat{M}_{i+1}^z+\frac{|\mf{s}|}{2}\right)\\
        &\hat{D}_i^3=\frac{2V^2}{q^3}\hat{L}_i^-\hat{L}_{i+1}^-\hat{M}_i^+\hat{M}_{i+1}^+\\
        &-\frac{2V^2}{q^3}\left(\hat{L}_i^z+\frac{q-|\mf{s}|}{2}\right)\left(\hat{L}_{i+1}^z+\frac{q-|\mf{s}|}{2}\right)\left(\hat{M}_i^z-\frac{|\mf{s}|}{2}\right)\left(\hat{M}_{i+1}^z-\frac{|\mf{s}|}{2}\right),
    \end{split}
\end{equation} 
where $\hat{J}_i^\alpha\coloneqq\hat{M}_i^\alpha+\hat{L}_i^\alpha$ is the sum of the $SU(1,1)$ spin operators.

\section{The discrete irrep of $SU(1,1)$}
\label{appsec:irrep of su(1,1)}
The $\mf{su}(1,1)$ algebra is defined by the following relations for the generators $\hat{K}^z$ and $\hat{K}^\pm$,
\begin{equation}
    [\hat{K}^z,\hat{K}^\pm]=\pm\hat{K}^\pm,\,\,[\hat{K}^+,\hat{K}^-]=-2\hat{K}^z.
\end{equation} 

The Casimir operator is defined as 
\begin{equation}
    \hat{C}\coloneqq(\hat{K}^z)^2-\frac{1}{2}\left(\hat{K}^+\hat{K}^-+\hat{K}^-\hat{K}^+\right).
\end{equation}

Just like $SU(2)$, irreps are characterized by the eigenvalues of $\hat{C}$ and $\hat{K}^z$, 
\begin{equation}
    \hat{C}\ket{k,\mu}=k(k-1)\ket{k,\mu},\,\,\hat{K}^z\ket{k,\mu}=(k+\mu)\ket{k,\mu}.
\end{equation}
The number $k$ is called the Bragmann index which is an analog of the spin quantum number for the $SU(2)$ spin. However, unlike the $SU(2)$ spin, $k$ can be any real number. Among possible combinations of $k$ and $\mu$, we consider the so-called (positive) discrete series, where $k$ is non-negative and $\mu$ is a non-negative integer. Due to non-compactness of $SU(1,1)$, the corresponding unitary irrep $D_k$ is infinite dimensional,
\begin{equation}
    D_k\coloneqq\mr{span}\left\{\big(\hat{K}^+\big)^n\ket*{k,\mu};\,n\in\mb{Z}_{\geq0}\right\}.
\end{equation} 

The coherent state $\ket{\bm{k}}=\ket{k\bm{n}}$ is parametrized by a unit psuedo-Eulcidean vector in the upper sheet of the hyperboloid $\mb{H}$,
\begin{equation}
    \bm{n}=\begin{pmatrix}
        \sinh\theta\cos\phi\\ 
        \sinh\theta\sin\phi\\
        \cosh\theta
    \end{pmatrix},\,\,(n^z)^2-((n^x)^2+(n^y)^2)=1,\,n^z>0,
\end{equation}
and is defined as 
\begin{equation}\label{eq:coherent state def}
    \ket{\bm{k}}\coloneqq\exp\left[\xi\hat{K}^+-\xi^*\hat{K}^-\right]\ket{k,0},\,\,\xi\coloneqq-\frac{\theta}{2}e^{-i\phi}.
\end{equation}
We briefly summarize basic properties of the coherent state necessary for the analysis of the $SU(1,1)$ spin dynamics.
\begin{itemize}
    \item The expectation value of the spin operators with respect to the coherent state is
    \begin{equation}
    \begin{split}
        \expval{\hat{K}^z}{\bm{k}}=k\cosh\theta,\,\expval{\hat{K}^\pm}{\bm{k}}=-ke^{\pm i\phi}\sinh\theta\,\,(\theta>0).
    \end{split}    
\end{equation}
    \item If the Bragmann index $k$ is greater than $1/2$, the coherent state is overcomplete and the resolution of the identity can be defined,
\begin{equation}\label{eq:resolution of identity}
    1=\int d\bm{k}\dyad{\bm{k}},
\end{equation}
where $d\bm{k}$ is the analog of the solid angle in $SU(2)$,
\begin{equation}
    d\bm{k}\coloneqq\frac{2k-1}{2\pi}\sinh\theta d\theta d\phi.
\end{equation}
    \item The overlap between two coherent states $\ket{\bm{k}}=\ket{k\bm{n}}$ and $\ket{\bm{k}'}=\ket{k\bm{n}'}$ is
    \begin{equation}\label{eq:overlap}
        \innerproduct*{\bm{k}'}{\bm{k}}=\left(\cosh\frac{\theta}{2}\cosh\frac{\theta'}{2}-\sinh\frac{\theta}{2}\sinh\frac{\theta'}{2}e^{i(\phi'-\phi)}\right)^{-2k},
    \end{equation}
    where $(\theta,\phi)$ and $(\theta',\phi')$ parametrize $\bm{n}$ and $\bm{n}'$, respectively.
\end{itemize}
Eq.\,\eqref{eq:resolution of identity} and Eq.\,\eqref{eq:overlap} imply that the topological term in the action $S_{\mr{top}}$ is written as,
\begin{equation}
    S_{\mr{top}}=ik\int dt\Big(\cosh\theta(t)-1\Big)\dv{\phi(t)}{t}.
\end{equation}

\section{The relevant irreps}
\label{appsec:relevant irreps}
In this section we provide a sketch of the proof that the physically relevant (local) subspace of $SU(1,1)^{\otimes2}$ should be totally symmetric with respect to permutations of flavors within $\mf{s}$ or its complement $\mf{q}\setminus\mf{s}$. The generators of $SU(1,1)^{\otimes2}$ are expressed as a sum of the following operators,
\begin{equation}
    \hat{m}_\mu^z\coloneqq\frac{1}{2}\left(\hat{a}_{\mu+}^\dag\hat{a}_{\mu+}+\hat{a}_{\mu-}^\dag\hat{a}_{\mu-}+1\right),\,\,\hat{m}_{\mu}^+\coloneqq\hat{a}_{\mu+}^\dag\hat{a}_{\mu-}^\dag,\,\,\hat{m}_{\mu}^-\coloneqq\hat{a}_{\mu-}\hat{a}_{\mu+},
\end{equation}
where we omit the site index. Due to the gauge invariance (or the particle number conservation) Eq.\,\eqref{eq:Schwinger}, while the irrep itself is infinite dimensional, the relevant physical subspace must be finite dimensional. This motivates us to define the following ``relevant'' subspace $S_{\mu}$,
\begin{equation}
    S_{\mu}\coloneqq\mr{span}\left\{\big(\hat{m}_{\mu}^+\big)^n\ket*{0};\,0\leq n\leq \kappa q\right\}.
\end{equation}
Therefore, the physical state must locally belong to $\bigotimes_{\mu\in\mf{s}}S_{\mu}\bigotimes_{\nu\not\in\mf{s}}S_{\nu}$. Since all the subspaces $S_{\mu}$ are isomorphic to each other, we can denote it by $V$, i.e., $S_{\mu}\cong S_{\nu}\cong V$. This implies that at each site the physical subspace is isomorphic to $V^{\otimes|\mf{s}|}\otimes V^{\otimes(q-|\mf{s}|)}$. Since $V$ is finite dimensional, we can apply the Schur-Weyl duality to each part ($V^{\otimes|\mf{s}|}$ and $V^{\otimes(q-|\mf{s}|)}$): $V^{\otimes|\mf{s}|}$ is decomposed as,
\begin{equation}
    V^{\otimes|\mf{s}|}\cong\bigoplus_\lambda\big(\mb{S}_\lambda(V)\big)^{\oplus m_\lambda},
\end{equation}
where $\mb{S}_\lambda(V)$ is the Weyl module corresponding to the Young tableau $\lambda$ (consisting of $|\mf{s}|$ boxes), and $m_\lambda$ is the multiplicity (see any standard textbook of the representation theory, e.g., Ref.\,\cite{Fulton2004} for the Schur-Weyl duality). Note that $D_{|\mf{s}|/2}\cap V_{\mr{phys}}$ belongs to $\lambda=(|\mf{s}|)$. The Weyl module is constructed by applying the Young symmetrizer from the right, and for any irrep corresponding to the tableau $\lambda\not=(|\mf{s}|)$, i.e., not totally symmetric, the Young symmetrizer contains the anti-symmetrizer. This implies that an arbitrary state in any other summand $\mb{S}_\lambda(V)$ is given as a superposition of states of the following form,
\begin{equation}\label{eq:AS|psi>}
    \widehat{AS}_{\mu\nu}\ket{n_1,\cdots,n_\mu,\cdots,n_\nu,\cdots,n_{|\mr{s}|}}\in\mb{S}_\lambda(V),\,\,\lambda\not=(|\mf{s}|),
\end{equation} 
where $\ket{n_1,\cdots,n_\mu,\cdots,n_\nu,\cdots,n_{|\mf{s}|}}$ is the Fock state in the number basis,
\begin{equation}
    \ket{n_1,\cdots,n_\mu,\cdots,n_\nu,\cdots,n_{|\mf{s}|}}\coloneqq\prod_{\gamma=1}^{|\mf{s}|}\frac{(\hat{m}_\gamma^+)^{n_\gamma}}{n_\gamma!}\ket{0}.
\end{equation}
$\widehat{AS}_{\mu\nu}$ is the projector onto the antisymmetric part of the wavefunction with respect to $\mu$ and $\nu$. Eq.\,\eqref{eq:AS|psi>} is written as 
\begin{equation}\begin{split}
    &\widehat{AS}_{\mu\nu}\ket{n_1,\cdots,n_\mu,\cdots,n_\nu,\cdots,n_{|\mr{s}|}}\\
    &=\frac{1}{2n_\mu!n_\nu!}\Big((\hat{m}_\mu^+)^{n_\mu}(\hat{m}_\nu^+)^{n_\nu}-(\hat{m}_\mu^+)^{n_nu}(\hat{m}_\nu^+)^{n_\mu}\Big)\ket{n_1,\cdots,0,\cdots,0,\cdots,n_{|\mr{s}|}}.
\end{split}
\end{equation}
By the inverse Choi–Jamio\l kowski isomorphism, this state is mapped to the following density matrix,
\begin{equation}
    \hat{\rho}=\frac{1}{2n_\mu!n_\nu!}\left((\hat{a}_\mu^\dag)^{n_\mu}(\hat{a}^\dag_\nu)^{n_\nu}\hat{\rho}'(\hat{a}_\nu)^{n_\nu}(\hat{a}_\mu)^{n_\mu}-(\hat{a}_\mu^\dag)^{n_\nu}(\hat{a}^\dag_\nu)^{n_\mu}\hat{\rho}'(\hat{a}_\nu)^{n_\mu}(\hat{a}_\mu)^{n_\nu}\right),
\end{equation}
where $\hat{\rho}'$ is the density matrix without $\hat{a}_\mu$ and $\hat{a}_\nu$ modes,
\begin{equation}
    \hat{\rho}'\coloneqq\left(\prod_{\gamma\not=\mu,\nu}\frac{1}{n_\gamma!}\right)\prod_{\gamma\not=\mu,\nu}(\hat{a}^\dag_\gamma)^{n_\gamma}\dyad*{0}\prod_{\gamma\not=\mu,\nu}(\hat{a}_\gamma)^{n_\gamma}.
\end{equation}
Since this state $\hat{\rho}$ is traceless, i.e., $\mr{tr}\hat{\rho}=0$, the irrep $\mb{S}_\lambda(V)$ for $\lambda\not=(|\mf{s}|)$ should not correspond to the physically relevant state. We repeat the same argument for the other part $V^{\otimes(q-|\mf{s}|)}$ to complete the proof of the statement.

\section{Another justification for the irrep}
\label{appsec:justification}
In this section we further justify our choice of the irrep $D_{|\mf{s}|/2}\otimes D_{(q-|\mf{s}|)/2}$ for the dynamics based on a simple mean-field analysis.

From Eq.\,\eqref{eq:su(1,1)+2}, we decompose the $SU(1,1)$ generators into their smaller components,
\begin{equation}
    \hat{m}_{i\mu}^z\coloneqq\frac{1}{2}\left(\hat{a}_{i\mu+}^\dag\hat{a}_{i\mu+}+\hat{a}_{i\mu-}^\dag\hat{a}_{i\mu-}+1\right),\,\,\hat{m}_{i\mu}^+\coloneqq\hat{a}_{i\mu+}^\dag\hat{a}_{i\mu-}^\dag,\,\,\hat{m}_{i\mu}^-\coloneqq\big(\hat{m}_{i\mu}^+\big)^\dag.
\end{equation}
These operators again form the $\mf{su}(1,1)$ algebra, and constitute the Hamiltonian $\hat{H}_{\mr{dis}}$. Unlike the $SU(1,1)$ generators $\hat{M}_i^\alpha$ and $\hat{L}_i^\alpha$, the initial state $\ket*{1_{SU(q)}}$ is an eigenstate of the Casimir operator with the Bragmann index $k=1/2$ for each $\hat{m}_{i\mu}^\alpha$, which implies that the state should belong to the tensor product of copies of this irrep,
\begin{equation}
    \ket*{\rho(t)}\in\bigotimes_{i=1}^L\bigotimes_{\mu=1}^qD_{1/2}.
\end{equation}
Therefore, it is justified to construct an approximate wavefunction within this Hilbert space. A natural choice for the mean-field ansatz is,
\begin{equation}\label{eq:ansatz appendix}
    \bigotimes_{i=1}^L\bigotimes_{\mu\in\mf{s}}\ket*{\bm{m}^{1/2}_i}_i\bigotimes_{\mu\not\in\mf{s}}\ket*{\bm{l}^{1/2}_i}_i,
\end{equation}
where $\ket*{\bm{m}^{1/2}_i}$ and $\ket*{\bm{l}^{1/2}_i}$ are the $SU(1,1)$ coherent states with $k=1/2$. This ansatz is invariant under permutations of indices within $\mf{s}$ and $\mf{q}\setminus\mf{s}$. Each factor can be written as,
\begin{equation}\begin{split}
    \bigotimes_{\mu\in\mf{s}}\ket*{\bm{m}^{1/2}_i}_i&=\exp\left[\xi_{i,m}\sum_{\mu\in\mf{s}}\hat{m}_{i\mu}^+-\xi^*_{i,m}\sum_{\mu\in\mf{s}}\hat{m}_{i\mu}^-\right]\ket*{\mr{vac}}=\exp\left[\xi_{i,m}\hat{M}_i^+-\xi^*_{i,m}\hat{M}_i^-\right]\ket*{\mr{vac}}\\
    \bigotimes_{\mu\not\in\mf{s}}\ket*{\bm{l}^{1/2}_i}_i&=\exp\left[\xi_{i,l}\sum_{\mu\not\in\mf{s}}\hat{m}_{i\mu}^+-\xi^*_{i,l}\sum_{\mu\not\in\mf{s}}\hat{m}_{i\mu}^-\right]\ket*{\mr{vac}}=\exp\left[\xi_{i,l}\hat{L}_i^+-\xi^*_{i,l}\hat{L}_i^-\right]\ket*{\mr{vac}},
    \end{split}
\end{equation}
where $\xi_{m,i}$ and $\xi_{l,i}$ are the corresponding complex numbers (see Eq.\,\eqref{eq:coherent state def}), and $\ket*{\mr{vac}}$ is the vacuum state in the bosonic Fock space. Since $\ket*{\mr{vac}}$ is an eigenstate of the Casimir operator for each $\hat{M}^\alpha_i$ and $\hat{L}_i^\alpha$ with the eigenvalue $|\mf{s}|/2$ and $(q-|\mf{s}|)/2$, respectively, the mean-field ansatz Eq.\,\eqref{eq:ansatz appendix} belongs to $D_{|\mf{s}|/2}\otimes D_{(q-|\mf{s}|)/2}$.

\section{Energy functional for the effective field theory}
The energy functional is constructed as the expectation value of the Hamiltonian $\hat{H}_{\mr{dis}}$ with respect to the coherent state. For each term $\hat{D}_i^\alpha$ defined in Eq.\,\eqref{eq:D su(1,1)}, in the large-$q$ limit, it is written as,
\begin{small}
\begin{equation}
    \begin{split}
        D_i^1&=\frac{qJ^2}{8}(1-r)^2\left|r\left(e^{i\Delta\phi_i}\sinh\theta_i\sinh\Theta_{i+1}+e^{i\Delta\phi_{i+1}}\sinh\Theta_i\sinh\theta_{i+1}\right)+(1-r)\sinh\Theta_i\sinh\Theta_{i+1}\right|^2\\
        &-\frac{qJ^2}{8}\left(4\kappa^2-r^2\left(\cosh\theta_i-1\right)\left(\cosh\theta_{i+1}-1\right)\right)\left(4(1+\kappa)^2-r^2\left(\cosh\theta_i+1\right)\left(\cosh\theta_{i+1}+1\right)\right)\\
        D_i^2&=\frac{qV^2}{8}r^2(1-r)^2\sinh\theta_i\sinh\theta_{i+1}\sinh\Theta_i\sinh\Theta_{i+1}e^{-i(\Delta\phi_i+\Delta\phi_{i+1})}\\
        &-\frac{qV^2}{8}r^2(1-r)^2(\cosh\theta_i+1)(\cosh\theta_{i+1}+1)(\cosh\Theta_i-1)(\cosh\Theta_{i+1}-1)\\
        D_i^3&=\frac{qV^2}{8}r^2(1-r)^2\sinh\theta_i\sinh\theta_{i+1}\sinh\Theta_i\sinh\Theta_{i+1}e^{i(\Delta\phi_i+\Delta\phi_{i+1})}\\
        &-\frac{qV^2}{8}r^2(1-r)^2(\cosh\theta_i-1)(\cosh\theta_{i+1}-1)(\cosh\Theta_i+1)(\cosh\Theta_{i+1}+1),
    \end{split}
\end{equation}
\end{small}
where $\Delta\phi_i\coloneqq\phi_i-\Phi_i$ is the relative phase. 
These terms become simplified for the spatially homogenous ansatz,
\begin{equation}\label{eq:D mf}
    \begin{split}
        D^1&=\frac{qJ^2}{8}(1-r)^2\Big|2re^{i\Delta\phi}\sinh\theta\sinh\Theta+(1-r)\sinh^2\Theta\Big|^2\\
        &-\frac{qJ^2}{8}\Big(4\kappa^2-r^2(\cosh\theta-1)^2\Big)\Big(4(1+\kappa)^2-r^2(\cosh\theta+1)^2\Big)\\
        D^2&=\frac{qV^2}{8}r^2(1-r)^2\left(e^{-2i\Delta\phi}\sinh^2\Theta\sinh^2\theta-(\cosh\Theta-1)^2(\cosh\theta+1)^2\right)\\
        D^3&=\frac{qV^2}{8}r^2(1-r)^2\left(e^{2i\Delta\phi}\sinh^2\Theta\sinh^2\theta-(\cosh\Theta+1)^2(\cosh\theta-1)^2\right).
    \end{split}
\end{equation}
The above expression implies that the energy functional is maximized when the phase difference is zero, $\Delta\phi=0$. By the construction of the coherent states, the angles $\theta$ and $\Theta$ are defined in $[0,\infty)$. Furthermore, due to the gauge invariance Eq.\,\eqref{eq:particle number each site}, $\cosh\Theta$ is a function of $\cosh\theta$. These conditions yield the energy functional as a function of the single variable, $\cosh\theta$, with the range specified in the main text.

\section{Details of the random phase approximation}
\label{appsec:RPA}
In this section we elaborate the stability analysis of Sec.\,\ref{sec:eft} within the random phase approximation (RPA). The small fluctuations around the mean-field value is represented by the variables $(\varphi,\vartheta)$, i.e., 
\begin{equation}
    \Delta\phi_i=0+\varphi_i,\,\,\theta_i=\theta+\vartheta_i,
\end{equation}
where $\theta$ is the mean-field value.
The topological term in the action is expanded as 
\begin{equation}
    S_{\mr{top}}[\varphi,\vartheta]=i\frac{|\mf{s}|}{2}\int d\tau\sum_{i=1}^L\sinh\theta\vartheta_i\dv{\varphi_i}{\tau},
\end{equation}
up to an irrelevant constant.
The dynamical part is similarly expanded as 
\begin{equation}
    \begin{split}
        S_{\mr{dyn}}[\varphi,\vartheta]&=S_{\mr{quad}}[\varphi,\vartheta]+S_J[\vartheta]\\
        S_{\mr{quad}}[\varphi,\vartheta]&=-q\int d\tau\sum_{i=1}^L\Big(t_\vartheta\vartheta_i\vartheta_{i+1}+m_\vartheta\vartheta^2_i+t_\varphi\varphi_i\varphi_{i+1}+m_\varphi\varphi_i^2\Big)\\
        S_J[\vartheta]&=q\int d\tau\sum_{i=1}^LJ_\vartheta\vartheta_i,
    \end{split}
\end{equation}
where the hopping parameters $t_\vartheta,t_\varphi$, the masses $m_\vartheta,m_\varphi$, and the ``source'' $J_\vartheta$ are functions of the mean-field value $\theta$. The quadratic part of this action $S_{\mr{quad}}[\varphi,\vartheta]$ is diagonal in the momentum space,
\begin{equation}
    S_{\mr{quad}}[\varphi,\vartheta]=-q\int d\tau\sum_{k=-\pi}^\pi\left[\Big(t_\vartheta\cos k+m_\vartheta\Big)\vartheta_{-k}\vartheta_k+\Big(t_\varphi\cos k+m_\varphi\Big)\varphi_{-k}\varphi_k\right].
\end{equation}

$t_\varphi$ and $m_\varphi$ are written as,
\begin{equation}
    \begin{split}
        t_\varphi&=\frac{V^2-J^2}{4}r^2(1-r)^2\sinh^2\theta\sinh^2\Theta\\
        m_\varphi&=\frac{1}{4}\Big((J^2+V^2)r^2(1-r)^2\sinh^2\theta\sinh^2\Theta+J^2r(1-r)^3\sinh\theta\sinh^3\Theta\Big).
    \end{split}
\end{equation}
One can find $t_\varphi<m_\varphi$ and $0<m_\varphi$ for $\forall\theta$. Therefore $t_\varphi\cos k+m_\varphi$ is always positive for $\forall\theta$ and for $\forall k\in[-\pi,\pi]$.

$t_\vartheta$ and $m_\vartheta$ are more complicated: we decompose each parameter into contributions from the unperturbed term $D_i^1$ and those from $D_i^2+D_i^3$, denoted by $t_J, m_J$ and $t_V, m_V$, respectively. $t_J$ and $m_J$ are defined as,
\begin{small}
\begin{equation}
    \begin{split}
        t_J&=-\frac{J^2}{4}(1-r)^2\left(r\left(\cosh\theta\sinh\Theta+\dv{\Theta}{\theta}\sinh\theta\cosh\Theta\right)+(1-r)\dv{\Theta}{\theta}\sinh\Theta\cosh\Theta\right)^2\\
        &-\frac{J^2}{4}(1-r)^2\dv{\Theta}{\theta}\sinh\Theta\cosh\Theta\left(2r\sinh\theta+(1-r)\sinh\Theta\right)\left(2r\cosh\theta+(1-r)\dv{\Theta}{\theta}\cosh\Theta\right)\\
        &-\frac{J^2}{4}r^2\sinh^2\theta\left(2(2\kappa^2+2\kappa+1)-r^2\left(\cosh^2\theta+1\right)\right)+\frac{J^2}{4}r^4\sinh^4\theta\\
        m_J&=-\frac{J^2}{4}(1-r)^2\left(r\Big(\cosh\theta\sinh\Theta+\dv{\Theta}{\theta}\cosh\Theta\sinh\theta\Big)+(1-r)\dv{\Theta}{\theta}\cosh\Theta\sinh\Theta\right)^2\\
        &-\frac{J^2}{4}(1-r)^2\sinh\Theta\left(r\sinh\theta\left(\left(1+\left(\dv{\Theta}{\theta}\right)^2\right)\sinh\Theta+\dv[2]{\Theta}{\theta}\cosh\Theta\right)\right.\\
        &\left.+(1-r)\sinh\Theta\left(\left(\dv{\Theta}{\theta}\right)^2\sinh\Theta+\dv[2]{\Theta}{\theta}\cosh\Theta\right)\right)(2r\sinh\theta+(1-r)\sinh\Theta)\\
        &-\frac{J^2}{4}r^2\cosh\theta\left(2(2\kappa^2+2\kappa+1)\cosh\theta-2(2\kappa+1)-r^2\sinh^2\theta\cosh\theta\right)\\
        &+\frac{J^2}{4}r^4\sinh^4\theta.
    \end{split}
\end{equation}
\end{small}
Note that $\Theta$ is a function of $\theta$ satisfying the gauge invariance Eq.\,\eqref{eq:particle number each site}. Therefore $\Theta$ should satisfy 
\begin{equation}
    \dv{\Theta}{\theta}=-\frac{r}{1-r}\frac{\sinh\theta}{\sinh\Theta},\,\,\dv[2]{\Theta}{\theta}=-\frac{r}{1-r}\left(\frac{\cosh\theta}{\sinh\Theta}+\frac{r}{1-r}\frac{\sinh^2\theta\cosh\Theta}{\sinh^3\Theta}\right).
\end{equation}

$t_V$ and $m_V$ are defined as,
\begin{small}
\begin{equation}
    \begin{split}
        t_V(\theta)&=-\frac{V^2}{4}r^2(1-r)^2\left(\cosh^2\theta\sinh^2\Theta+\left(\dv{\Theta}{\theta}\right)^2\sinh^2\theta\cosh^2\Theta\right)\\
        &-\frac{V^2}{2}r^2(1-r)^2\dv{\Theta}{\theta}\sinh\theta\cosh\theta\sinh\Theta\cosh\Theta\\
        &+\frac{V^2}{4}r^2(1-r)^2\left(\sinh^2\theta\left(\cosh^2\Theta+1\right)+\left(\dv{\Theta}{\theta}\right)^2\sinh^2\Theta\left(\cosh^2\theta+1\right)\right)\\
        &+\frac{V^2}{2}r^2(1-r)^2\dv{\Theta}{\theta}\sinh\theta\sinh\Theta(\cosh\theta\cosh\Theta-1)\\
        m_V(\theta)&=-\frac{V^2}{4}r^2(1-r)^2\sinh^2\theta\sinh\Theta\left(\left(1+\left(\dv{\Theta}{\theta}\right)^2\right)\sinh\Theta+\dv[2]{\Theta}{\theta}\cosh\Theta\right)\\
        &-\frac{V^2}{4}r^2(1-r)^2\dv{\Theta}{\theta}\sinh\theta\cosh\theta\sinh\Theta\cosh\Theta\\
        &+\frac{V^2}{4}r^2(1-r)^2\cosh\theta\left(\cosh\theta(\cosh^2\Theta+1)-2\cosh\Theta\right)\\
        &+\frac{V^2}{4}r^2(1-r)^2\left(\dv[2]{\Theta}{\theta}\sinh\Theta+\left(\dv{\Theta}{\theta}\right)^2\cosh\Theta\right)\left(\cosh\Theta(\cosh^2\theta+1)-2\cosh\theta\right)\\
        &+\frac{V^2}{2}r^2(1-r)^2\dv{\Theta}{\theta}\sinh\theta\sinh\Theta(\cosh\theta\cosh\Theta-1).
    \end{split}
\end{equation}
\end{small}

\bibliographystyle{unsrt}
\bibliography{reference}
\end{document}